\documentclass[preprint,12pt,authoryear]{elsarticle}

\usepackage{amssymb}
\usepackage{amsmath}
\usepackage{graphicx}
\usepackage{natbib} 

\usepackage{url} 

\usepackage{subcaption}
\usepackage{graphicx}
\usepackage{dcolumn}
\usepackage{amsmath,amsfonts, dsfont}
\usepackage{bm}
\usepackage{hyperref}
\usepackage{epsfig,psfrag,color}
\usepackage{amsfonts, dsfont}
\usepackage{latexsym}
\usepackage{times}
\usepackage{natbib}
\usepackage{arydshln}
\usepackage{lineno}
\usepackage{setspace}
\usepackage{multicol}
\usepackage{multirow}
\usepackage{diagbox}

\newcommand{\spn}{\mathrm{span}}

\newcommand{\E}{\mathrm{E}}

\newcommand{\cov}{\mathrm{cov}}
\newcommand{\Cov}{\mathrm{Cov}}

\newcommand{\tr}{\mathrm{\,tr}}
\newcommand{\vecc}{\mathrm{\,vec}}

\newcommand{\real}{{\mathbb R}}
\newcommand{\ind}{\mathbb I}
\newcommand{\Wbf}{\mathbf W}
\newcommand{\cerobf}{\bm 0}
\newcommand{\ubf}{\mathbf u}

\newcommand{\Vbf}{\mathbf V}

\newcommand{\ones}{\bm 1}
\newcommand{\X}{\mathbf X}
\newcommand{\s}{\mathbf s}

\newcommand{\x}{{\mathbf x}}
\newcommand{\Y}{{\mathbf Y}}

\newcommand{\f}{{\mathbf f}}

\newcommand{\Ab}{\mathbf A}
\newcommand{\Bb}{\mathbf B}
\newcommand{\Cb}{\mathbf C}

\newcommand{\Hb}{\mathbf H}

\newcommand{\Sb}{\mathbf{S}}

\newcommand{\Xbb}{\mathbb{X}}
\newcommand{\Fbb}{\mathbb{F}}
\newcommand{\Xbbb}{\bar \Xbb}
\newcommand{\Fbbb}{\bar \Fbb}

\newcommand{\Zbb}{\mathbb Z}
\newcommand{\Ebb}{\mathbb{E}}
\newcommand{\Nbb}{\mathbb{N}}
\newcommand{\Vbb}{\mathbb{V}}
\newcommand{\Ubb}{\mathbb{U}}

\makeatletter
\renewcommand*{\@seccntformat}[1]{%
 \csname the#1\endcsname.\quad}
\makeatother

\newcommand{\Cbf}{{\mathbf C}}

\newcommand{\W}{{\mathbf W}}

\newcommand{\Hbf}{{\mathbf H}}

\newcommand{\greekbold}[1]{\mbox{\boldmath $#1$}}
\newcommand{\alphabf}{\greekbold{\alpha}}

\newcommand{\lambdabf}{\greekbold{\lambda}}

\newcommand{\epsilonbf}{\greekbold{\epsilon}}

\newcommand{\Deltabf}{\greekbold{\Delta}}

\newcommand{\Sigmabf}{\greekbold{\Sigma}}

\newcommand{\Sigmabfs}{{\greekbold{\scriptstyle \Sigma}}}
\newcommand{\mubf}{\greekbold{\mu}}

\newcommand{\thetabf}{\greekbold{\theta}}

\newcommand{\nubf}{\greekbold{\nu}}

\newcommand{\bspc}{\mathcal B}

\newcommand{\isebjhat}[1]{\widehat{{\cal IE}}_{\Sigmabfs}(\bspc)}

\newtheorem{theorem}{Theorem}[section]

\newtheorem{prop}[theorem]{Proposition}

\newenvironment{proofprop}[1]{\textbf{Proof of Proposition \ref{#1}:} }{
\begin{flushright}
\rule{0.5em}{0.5em}
\end{flushright}}

\begin{document}

\begin{frontmatter}
\title{Sufficient dimension reduction for regression with spatially correlated errors: application to prediction} 
\author[label1,label2]{Liliana Forzani}
\author[label2,label3]{Rodrigo Garc\'ia Arancibia} 
\author[label1]{Antonella Gieco}
\author[label1,label2]{Pamela Llop} 
\author[label4]{Anne Yao} 

\affiliation[label1]{organization={Facultad de Ingenieria Quimica, Universidad Nacional del Litoral },
            city={Santa Fe},
            postcode={3000}, 
            state={Santa Fe},
            country={Argentina}}
\affiliation[label2]{organization={Consejo Nacional de Invesitgaciones Cientificas y Tecnicas (CONICET)},
            city={Santa Fe},
            postcode={3000}, 
            state={Santa Fe},
            country={Argentina}}            
\affiliation[label3]{organization={Instituto de Economia Aplicada Litoral , Universidad Nacional del Litoral },
            city={Santa Fe},
            postcode={3000}, 
            state={Santa Fe},
            country={Argentina}}
\affiliation[label4]
 {organization={CNRS, Laboratoire de Mathématiques Blaise Pascal, Université Clermont Auvergne},
            city={Clermont-Ferrand},
            postcode={63100}, 
            state={Puy-de-Dome},
            country={France}}

\begin{abstract}

In this paper, we address the problem of predicting a response variable in the context of both, spatially correlated and high-dimensional data. To reduce the dimensionality of the predictor variables, we apply the sufficient dimension reduction (SDR) paradigm, which reduces the predictor space while retaining relevant information about the response. To achieve this, we impose two different spatial models on the inverse regression: the separable spatial covariance model (SSCM) and the spatial autoregressive error model (SEM). For these models, we derive maximum likelihood estimators for the reduction and use them to predict the response via nonparametric rules for forward regression. Through simulations and real data applications, we demonstrate the effectiveness of our approach for spatial data prediction.

\end{abstract}

\begin{highlights} 
\item We introduce a Sufficient Dimension Reduction (SDR) methodology  for spatial data.
\item Using a model-based inverse regression approach, two alternative models for spatially correlated errors are proposed: the Separable Spatial Covariance Model (SSCM) and the Spatial Autoregressive Error Model (SEM).
\item Maximum likelihood estimators for the dimension reduction are derived. 
\item Spatial nonparametric regression models are used to assess the predictive 
performance of the proposed methodology. 
\item Simulations and real data applications show the advantages of the spatial SDR approach for geostatistical and lattice data prediction
. \end{highlights}
\noindent%

\vfill
\end{frontmatter}

\newpage
\section{Introduction}

The prediction problem is currently one of the primary concerns in data science and statistical learning, driving the development of novel approaches in theoretical, computational, and applied statistics across a wide range of scientific disciplines. In particular, many real-world applications involve dependent data, such as time series, longitudinal measurements, or spatial data. It is well known that predictive models designed for independent data do not perform well in these scenarios, necessitating the development of specific methodologies to address these issues. More specifically, for spatial data modeling, it is appropriate to include the information about neighboring similarities provided by both, the distance between locations and the values of variables measured in such locations. The use of such information to predict a variable of interest in new (unsampled) locations is commonly called spatial prediction.  This issue was traditionally addressed using geostatistical models, from which the classical kriging and co-kriging predictors were derived. Today, the spatial prediction problem is also covered by spatial econometrics \citep[e.g.]{hu2009spatial, goulard2017, Kop23} and  machine learning approaches \citep[e.g.][]{ heaton2019case,meyer2021predicting, Nikparvar21}.

Formally, spatial prediction involves a  response (target) variable $Y_{\s}$ to predict in some unobserved location $\s_0 \in D\subset \real^2$ given a set of $p$ predictor variables $\X_{\s_0}=(X_{\s_0,1},X_{\s_0,2},\ldots,X_{\s_0, p})^T$ observed in a location $\s_0$. For this purpose, we can consider the spatial regression model 
\begin{equation}\label{genmodel}
    Y_{\s} \mid \X_{\s} = \eta(\X_{\s})+\varepsilon_{\s},
\end{equation}
where $Y_{\s} \in \real$ is the response variable, $\X_{\s}=(X_{\s,1},X_{\s,2},\ldots,X_{\s, p})^T$ is a vector of $p$ continuous predictor variables measured in a location $\s \in D\subset \real^2$, $\eta:\real^p \longrightarrow \real$ is an unknown function,  $\varepsilon_{\s}$ represents a zero-mean error term. 
For the function $\eta(\cdot )$ we can assume a spatial non-parametric model as in \citep{Biau2004,DaboNiang2007,menezes10}, given the advantages derived from a flexible model for spatial prediction. However, if the number of predictor variables ($p$) is large, we face the well-known {\it curse of dimensionality} inherent in such models. To leverage the benefits of non-parametric modeling for spatial data, we propose reducing the dimensionality of the predictors by exploiting all the information contained in the data, and then applying non-parametric regression as the predictive method.

Specifically, our aim is to find a lower dimensional function of the predictor vector $\X_{\s}$, $R: \real^p \longrightarrow \real^d$ with $d \leq p$, that contains all relevant information for the response $Y$, that is, such that $Y_{\s}\mid \X_{\s} \, \buildrel d \over = \, Y_{\s}\mid R(\X_{\s})$, where $\buildrel d \over =$ means equal in distribution. 
Here the function $R(.)$ constitutes the so-called {\it Sufficient Dimension Reduction} (SDR) of the regression of $Y_{\s}$ on $\X_{\s}$. Finding the SDR is surprisingly straightforward when using the method of \textit{inverse regression}. Inverse regression finds its foundation in  the equivalence \citep{cook98, Cook2007}
\begin{equation}\label{equivalencia}
Y_{\s}\mid \X_{\s} {\,\,  \buildrel d \over = \,\,  } Y_{\s}|R(\X_{\s})  \quad \Longleftrightarrow \quad \X_{\s} \mid (Y_{\s}, R(\X_{\s})) {\,\, \buildrel d \over = \,\, } \X_{\s}\mid R(\X_{\s}).
\end{equation}
Observe that statement in the left-hand side defines, as stated before, the sufficient reduction $R(\X_{\s})$ for the forward regression \eqref{genmodel} whereas the right-hand side defines the sufficient statistic $R(\X_{\s})$ for the \textit{parameter} $Y_{\s}$ of the inverse regression $\X_{\s}\mid Y_{\s}$. Equivalence \eqref{equivalencia} then states that, if $Y_{\s}$ is considered as the parameter of the model, the sufficient statistic for $Y_{\s}$ is the sufficient reduction for the regression $Y_{\s} \mid \X_{\s}$. In this context, solving the inverse problem of finding a sufficient statistic for the regression of $\X_{\s}$ on $Y_{\s}$ allows us to find the sufficient reduction for $Y_{\s}$ on $\X_{\s}$. When the model imposed for regression of $\X_{\s}$ on $Y_{\s}$ allows us to specify a sufficient statistic for $Y_{\s}$, then the reduction is \textit{minimal} sufficient and, in this case, the sufficient dimension reduction approach is called \textit{model-based}. 

Following the model-based inverse regression approach, to find the reduction $R(\cdot)$ it is not necessary to assume a specific distribution for the forward regression $Y_{\s} \mid \X_{\s}$. In this context, after finding the reduction, any regression function can be used to model such regression. In particular, for the spatial context, model \eqref{genmodel} would change to 
\begin{align}\label{genmodel-2}
    Y_{\s} \mid \X_{\s} &= {\eta}(\X_{\s})+\varepsilon_{\s} \nonumber\\
    &= \widetilde{\eta}(R(\X_{\s}))+\varepsilon_{\s},
\end{align}
with $\widetilde{\eta}:\real^d \longrightarrow \real$ and $d\leq p$ a new, lower dimensional based, regression function.

For independent data, \textit{model-based} SDR techniques for continuous predictors were introduced and studied in \cite{Cook2007, CookForzani2008,BDF16} and for mixtures of binary, ordinal and continuous predictors, in \cite{duarte2023,forzani18,bura22}. In this context, \cite{forzani2024asymptotic} obtained asymptotic results for the non-parametric regression estimator of $\tilde{\eta}$ regardless of whether the true $R(.)$ or its estimator is used. Another SDR approaches are the \textit{moment-based} ones, such as the \textit{Sliced Inverse Regression} (SIR), \citep{Li1991}, the \textit{Sliced Average Variance Estimation} (SAVE) \citep{CookWeisberg1991}, and the  \textit{Directional Regression} (DR) \citep{LiWang2007}, but in general they are not exhaustive. Finally, another SDR technique in the independent context is the \textit{Principal Support Vector Machines} (SVM) \cite{Li1991}, for binary response. 

Although all these methods were introduced for independent data, some of them have been extended to the dependent context, both spatially correlated and time series data. In a time series context,  \cite{BARBARINO20241,Barbarino2015} analyzed the applicability of SIR for independent data to forecasting and extended it to time-series data (TSIR), showing its consistency for covariance-stationary predictor time series. \cite{matilainen2019sliced} made a similar contribution by extending SAVE.

Recently, various dimension reduction methods for spatial data have been proposed under different approaches. These include a generalization of the SIR under some mixing conditions \cite{loubes13}, partial least squares \citep{SAMPSON2013383}, principal component regression \citep{junttila2017bayesian}, and envelope methods \citep{MAY2024100838}. In the point processes setting, an SDR paradigm has been introduced by \cite{guan08,guan10}, which allows to extend most of the popular inverse regression techniques to this context. More recently, \cite{datta2022} extended the SVM methodology to the point pattern process context based on the idea of \textit{Central Intensity Subspace}.  However, under the model-based SDR paradigm, there still exists a significant gap that this work aims to fill. 

In this paper, we propose a sufficient dimension reduction $R(\cdot)$ for spatial data and apply it to perform prediction using nonparametric rules for $\widetilde{\eta}(\cdot)$ in \eqref{genmodel-2}. Moreover,  we consider the predictor  proposed by \cite{DN16}, where the distance between locations is incorporated into the weights of the nonparametric regression  $\widetilde{\eta}(\cdot)$. Based on the inverse regression approach, we present two models with alternative structures for spatial dependency in the error terms, applicable to both geostatistical and lattice data within a fixed domain $D$. The first error model specifies a separable cross-covariance matrix that controls spatial association using a function dependent on the distance between spatial coordinates $\s \in D$. The second model, more aligned with the spatial econometrics approach, captures the spatial dependence of the error terms through an autoregressive structure, employing a spatial weights matrix that specifies neighbors for each observation located at $\s \in D$. Using a more general formulation that encompasses both models, we get the sufficient dimension reduction for the spatial regression of $Y_s$ on $\X_s$, deriving then the maximum likelihood estimators for each model.

The rest of this paper is organized as follows. In Section \ref{secInvModelSDR} we present our general inverse regression model in the context of spatial data, deriving the sufficient dimension reduction for $Y_{\s}\mid \X_{\s}$. Sections \ref{secSSCM} and \ref{secSEM} focus on the specific spatial separable covariance (SSCM) and spatial autoregressive error (SEM)
models, as well as the maximum likelihood estimates of the sufficient dimension reductions. In Section \ref{prediction} we present the two non-parametric strategies that we adopted for spatial prediction using our dimension reduction methodologies. Section \ref{simulations} details a simulation experiment to compare the predictive performance of our spatial SDR models and their behavior under varying numbers of predictors and sample sizes. In Section \ref{realdata}, we analyze three real data applications to illustrate the practical utility of our approaches for spatial prediction as well as to show the predictive improvements obtained with our proposed methods. Then, in Section \ref{discussion}, we discuss the use and performance of our methodologies in the contexts of geostatistics and lattice data, based on the results from simulation and the real data applications. Finally, a brief conclusion is provided in Section \ref{conclus}.

\section{Spatial inverse regression model and sufficient dimension reduction}\label{secInvModelSDR}

For the regression of $Y$ on $\X$, where $Y \in \real$ and $\X \in \real^p$, in the context of independent data, \cite{Cook2007}, \cite{CookForzani2008}, and \cite{BDF16} demonstrate that for the inverse regression model

\begin{equation}\label{modelo1}
    \X|Y =\mubf_Y +\epsilonbf,
\end{equation} 
with $\epsilonbf \sim \mathcal{N}(\mathbf{0}, \Deltabf)$, if $\alphabf$ is a basis for the subspace $\mathcal{S_{\alpha}} = \Deltabf^{-1} \spn \{\mubf_Y-{\mubf}, Y\in \mathcal{S}_Y\}$, where $\mubf = E_Y(\mubf_Y)$ and $\mathcal{S}_Y$ denotes the sample space of $Y$, then $R(\X) = \alphabf^T(\X - E(\X))$ is the {\it minimal sufficient reduction} for the regression of $Y$ on $\X$. In this direction, if  $\Ab \in \real^{p\times d}$ with $d\leq p$  is a basis for the $\spn \{\mubf_Y-{\mubf}, Y\in \mathcal{S}_Y\}$, then $\mubf_Y-\mubf=\mathbf{A}\nubf_Y$ for some $\nubf_Y \in \real^d$ and therefore the model \eqref{modelo1} can be written as 
\begin{equation}\label{modelo2}
\X|Y=\mubf + \Ab\nubf_Y + \epsilonbf.
\end{equation}
In this model, the dependence of  $\nubf_Y$  on $Y$ could be modeled as a function $\f_{Y}\in\mathbb{R}^{r}$ of $Y$. Specifying for $\X|Y$ a linear model with predictor vector $\f_Y$, gives the so-called {\it Principal Fitted Component (PFC) models} of \cite{CookForzani2008}.  When $Y_{\s}$ is continuous, $\f_{Y}$
usually will be a flexible set of basis functions, like polynomial terms in $Y$, which may also be used when it is impractical to apply graphical methods to all of the predictors \citep{AC09}. When $Y$ is categorical and takes values $\{C_{1},\ldots,C_{h}\}$, we can
set $r=h-1$ and specify the $j$-th element of $\f_{Y}$ to be $\mathbb{I}(y\in C_{j})$, $j=1,\ldots,h$. We can also, when $Y$ is continuous,  \textit{slice} its values into $h$ categories $\{C_{1},\ldots,C_{h}\}$ and then specify the $j$-th coordinate of $\f_{Y}$ as for the
case of a categorical $Y_{\s}$ (for more details see \cite{AC09}). As a consequence, model \eqref{modelo1} can be written as 
\begin{equation}\label{model-last}
\X|Y= \mubf + \Ab\Bb \f_{Y}+\epsilonbf,
\end{equation}
and the sufficient reduction for the regression of $Y$ on $\X$ is given by $R(\X)=\Ab^T\Deltabf^{-1}\X$. 

\

In the context of spatial data, let us consider the spatial regression of a response $Y_{\s} \in \real$  from a second order stationary spatial process,  on a set of $p$ continuous predictor variables $\X_{\s}=(X_{\s,1},X_{\s,2},\ldots,X_{\s, p})^T$ measured in a location $\s \in D\subset \real^2$. In other words, we will consider a sample $\{\X_{\s_i}, Y_{\s_i}\}_{i=1}^n$ of predictor vectors $\X_{\s_i} \in \mathbb{R}^p$ and response variables $Y_{\s_i} \in \mathbb{R}$ measured in $n$ spatial locations $i=1,\ldots, n$. Defining $\Xbb \in \real^{n\times p}$ as the matrix such that $\Xbb^T \doteq (\X_{\s_1}, \ldots, \X_{\s_n})$, $\Y \in \real^n$ as the vector $\Y^T \doteq (Y_{\s_1},\dots, Y_{\s_n})$, and $\Ebb  \in {\real}^{n\times p}$ as $\Ebb^T \doteq (\epsilonbf_{\s_1}, \ldots, \epsilonbf_{\s_n})$, the matrix form of model \eqref{modelo1} is given by

\begin{equation}\label{spatial-model}
	\Xbb |\Y  =  \Nbb_Y + \Ebb,
\end{equation}
where $\Nbb_Y \in {\real}^{n\times p}$ is such that $\Nbb_Y^T \doteq (\mubf_{Y_{\s_1}}, \ldots, \mubf_{Y_{\s_n}})$. In this model, we will assume that the errors $\epsilonbf_{\s_i}$, $i = 1,\ldots,n$, {follow a second-order stationary process, are independent of $Y_{\s_i}$, and are normally distributed with mean $E(\epsilonbf_{s})=\cerobf$ and positive-definite and constant covariance matrix $\cov(\epsilonbf_{\s})=\Deltabf\in\real^{p\times p}$}.
In addition, for the spatial dependence of the error, for any pair $i,j$, we will consider the positive definite matrix $\Sb\in \real^{n\times n}$ of elements $\Sb_{ij}$ that  represents the spatial association between $\epsilonbf_{\s_i}$ and $\epsilonbf_{\s_j}$ 
via \textit{cross-covariance} or, \textit{spatial weight} matrix, as will be specified in the next section. 

In this way, the error has distribution is $\vecc(\Ebb^T) \sim \mathcal{N}(\cerobf_{np}, \Sb \otimes\Deltabf)$ so that the log-likelihood function $l(\Xbb|\Y;\Sb,\Deltabf) =\log  L(\Xbb|\Y;\Sb,\Deltabf)$  for  model \eqref{spatial-model} is given by 
\begin{align} \label{log-likelihood}
l(\Xbb|\Y;\Sb,\Deltabf) = - \frac{np}{2} \log(2\pi)- \frac{p}{2} \log | \Sb|  - \frac{n}{2} \log |\Deltabf|-\frac{1}{2} \tr\big((\Xbb  - \Nbb_Y) \Deltabf^{-1} (\Xbb  - \Nbb_Y)^T \Sb^{-1}\big). 
\end{align}
From this log-likelihood function, we will obtain the sufficient dimension reduction for the spatial regression of $\Y$ on $\Xbb$, as stated in following result.  
\begin{theorem}\label{teo1}
If $\Xbb | \Y$ has log-likelihood given by  \eqref{log-likelihood} , then a sufficient reduction for the regression of $\Y|\Xbb$  is given by
\begin{equation} \label{reduction}
R(\Xbb) =  \Xbb \Deltabf^{-1} \Ab,
\end{equation}
where $\Ab$ is a base for the $\spn\{\mubf_{{Y}_{\s_i}}-\mubf, {Y}_{\s_i} \in \mathcal{S}_Y \}$ with $\mubf=E(\mubf_{{Y}_{\s_i}})$ for all $i=1, \ldots, n$ 
\end{theorem}

\begin{proof} 
This proof is based in the ideas stated in \cite{Cook2007} (see also \cite{BDF16}) using the Lehmann-Scheff\'e Theorem  for sufficient statistics. In this direction, to prove the result, we will prove that if $R(\Xbb)$ is a sufficient statistics for $\Y$ (where $\Y$ is considered as a parameter) then $\Xbb \mid (\Y, R(\Xbb)) \sim \Xbb \mid R(\Xbb)$. Now, by the the equivalence \eqref{equivalencia} (see \cite{Cook2007}) it follows that $\Y\mid \Xbb \sim \Y|R(\Xbb)$, i.e., $R(\Xbb)$ is a sufficient reduction for the regression of $\Y$ on $\Xbb$. 

Then, by the  Lehmann-Scheff\'e Theorem (Theorem 6.2.13 in \cite{casellaberger90}) we have that, if for every two sample points $\Xbb$, $\Zbb$ the ratio $ L(\Xbb|\Y;\Sb,\Deltabf)/ L(\Zbb|\Y;\Sb,\Deltabf)$ is independent of $\Y$ if only if $R(\Xbb) = R(\Zbb)$ then $R(\Xbb)$ is sufficient for the regression of $\Y|\Xbb$. Equivalently, let us prove that $l(\Xbb|Y;\Sb,\Deltabf)-l(\Zbb|Y;\Sb,\Deltabf)$ is independent of $\Y$ if only if $R(\Xbb) = R(\Zbb)$. Except for some constants, from \eqref{log-likelihood} we have that
\[
l(\Xbb|\Y;\Sb,\Deltabf)-l(\Zbb|\Y;\Sb,\Deltabf) = \tr\Big( \Xbb\Deltabf^{-1} \Xbb^T \Sb^{-1} - \Zbb\Deltabf^{-1} \Zbb^T \Sb^{-1} - 
 2(\Xbb-\Zbb) \Deltabf^{-1}\Nbb_Y^T \Sb^{-1}\Big),
\]
which is independent of $\Y$ if, for some constant matrix $\Cbf \in \real^{n \times n}$, independent of $\Y$, it is verified that 
\[
\Xbb\Deltabf^{-1} \Xbb^T \Sb^{-1} - \Zbb\Deltabf^{-1} \Zbb^T \Sb^{-1} -  2(\Xbb-\Zbb) \Deltabf^{-1}\Nbb_Y^T \Sb^{-1} = \Cbf.
\]
Taking expectation with respect to $\Y$ it yields,
\[  (\Xbb-\Zbb) \Deltabf^{-1}(\Nbb_Y - E_Y(\Nbb_Y))^T\Sb^{-1} = \cerobf_{n\times n},
\]
or equivalently, 
\begin{equation}\label{esta}
  (\Xbb-\Zbb) \Deltabf^{-1}(\Nbb_Y - E_Y(\Nbb_Y))^T = \cerobf_{n\times n}.
\end{equation}
Since $\Ab \in \real^{p \times d}$ is a basis for $\text{span}\{\mubf_{Y_{\s_i}} - \mubf, Y_{\s_i} \in \mathcal{S}_Y, i= 1,\ldots,n \}$,  $\mubf_{Y_{\s_i}} - \mubf = \Ab \nubf_{Y_{\s_i}}$ for some $\nubf_{Y_{\s_i}} \in \real^d$. Defining $\Vbb \in {\real}^{n\times d}$ as $\Vbb_Y^T \doteq (\nubf_{Y_{\s_1}}, \ldots, \nubf_{Y_{\s_n}})$ we have that 
\begin{equation}\label{eqnyvy}
\Nbb_Y - E_Y(\Nbb_Y) = \Nbb_Y - \ones_n\mubf^T = \Vbb_Y \Ab^T,    
\end{equation}
 where $\ones_n$ will indicates the $n$-dimensional vector of ones. Finally, plug-in \eqref{eqnyvy} in \eqref{esta} we have that,
\[
(\Xbb-\Zbb) \Deltabf^{-1} \Ab \Vbb_Y^T  = \cerobf_{n\times n} \quad \Longleftrightarrow \quad \Xbb\Deltabf^{-1} \Ab = \Zbb\Deltabf^{-1} \Ab,
\]
then, by Lehmann-Scheff\'e Theorem $R(\Xbb) =  \Xbb \Deltabf^{-1} \Ab$ is sufficient for the regression $\Y|\Xbb$. 
$\hfill\square$
\end{proof}

As a consequence of \eqref{eqnyvy}, model \eqref{spatial-model} can be written as,
\begin{equation}
\Xbb |\Y = \ones_n  \mubf^T + \Vbb_Y \Ab^T + \Ebb.
\end{equation}
Following the ideas in \cite{CookForzani2008,AC09} stated in Model 
\eqref{model-last}, the dependence of the functions $\nubf_{Y_{\s_i}}$ on $Y_{\s_i}$ (and, as a consequence the dependence of $\mubf_{Y_{\s_i}}$ on $Y_{\s_i}$), for $i= 1,\ldots,n$, can be modeled using $r$ functions $\f_{Y_{\s_i}}\in \mathbb{R}^{r}$ so that, given the matrix of parameters $\Bb\in \mathbb{R}^{d\times r}$, $\Vbb = \Fbb\Bb^T$ with $\Fbb^T \in \mathbb{R}^{n\times r}$ as $\Fbb^T \doteq (\f_{Y_{\s_1}}, \dots, \f_{Y_{\s_n}})$. Finally, the spatial model is given by
\begin{equation}\label{spatial-model-1}
\Xbb |\Y  =  \ones_n  \mubf^T + \Fbb ( \Ab \Bb)^T + \Ebb,
\end{equation}
with $\vecc(\Ebb^T) \sim \mathcal{N}(\cerobf_{np}, \Sb \otimes\Deltabf)$. 

Subsequently, from theorem \ref{teo1} we establish that  $R(\Xbb) =  \Xbb \Deltabf^{-1} \Ab$ serves a sufficient dimension reduction for the regression or $\Y$ on $\Xbb$. Hence, we need to estimate $\Deltabf$ and $\Ab$, both of which rely on the specification of the spatial association matrix $\Sb$ for the errors.
In the following sections, we delineate two specific instances of the inverse Model \eqref{spatial-model-1}. Firstly, by employing $\Sb$ as a cross-covariance matrix, we engage with the Spatial Separable Covariance Model (SSCM) for the inverse regression. Secondly, opting for a weight matrix of spatial lags leads us to consider a Spatial Autoregressive Error Model (SEM). For both model we present maximun likelihood estimators of the sufficient dimension reduction.

\section{Spatial Separable Covariance Model (SSCM)} \label{secSSCM}

To state this model, in \eqref{spatial-model-1} we assume that the errors $\epsilonbf_{\s_i}$ for $i = 1,\ldots,n$,  has normal distribution with zero mean, covariance matrix $\cov(\epsilonbf_{\s_i})=\Deltabf \in \real^{p\times p}$ and \textit{cross-covariance} $\cov (\epsilonbf_{\s_i}, \epsilonbf_{\s_j}) = \rho (d(\s_i - \s_j)) \Deltabf$ for any pair $i,j$, where $\rho (x):{\real}^+_0 \rightarrow {\real}^+_0$ that controls the spatial association. If $\Hb\in \real^{n\times n}$  is the positive definite matrix   with elements, where $\Hb_{ij}= \rho (d(\s_i - \s_j))$. Then the SSCM model assumes that   $\vecc(\Ebb^T)  \sim \mathcal{N}(\cerobf_{np}, \Hb \otimes\Deltabf)$ with the cross-covariance $np \times np$ matrix $\Hb \otimes\Deltabf$ is given by 
\[ 
\Hb \otimes \Deltabf = \left( \begin{array}{cccc}
	\Deltabf & \Deltabf \rho (|\s_1-\s_2|) & \dots &  \Deltabf \rho (|\s_1-\s_n|)\\
	\dots & \dots & \dots & \dots \\
	\dots & \dots & \dots & \dots \\
	\Deltabf  \rho (|\s_1-\s_n|) & \Deltabf \rho (|\s_2-\s_n|) & \dots &  \Deltabf  
\end{array}\right),
\]
which is model \eqref{spatial-model-1} with $\Sb = \Hb$.  

From this model we can obtain the log-likelihood function of the data which is given by
\begin{align} \label{log-likelihood-SSCM}
	l(\Xbb|\Y;\mubf, \Ab, \Bb, \Deltabf,  \Hb) &= - \frac{np}{2} \log(2\pi)- \frac{p}{2} \log | \Hb|  - \frac{n}{2} \log |\Deltabf| \\ 
	&\hspace{-1cm} -\frac{1}{2} \tr \big( \Hb^{-1/2} (\Xbb -  \ones_n  \mubf^T - \Fbb \Bb^T \Ab^T) \Deltabf^{-1}  (\Xbb -  \ones_n  \mubf^T - \Fbb \Bb^T \Ab^T)^T  \Hb^{-1/2} \big). \nonumber
\end{align}
From this equation the maximum likelihood estimators for the parameters of the SSCM model, and consequently the estimated SDR, can be obtained. This is stated in the following result. 
\begin{prop}\label{SSCM-estimators}
Given the matrix $\Hb$, the maximum likelihood estimators
 of the parameters of the SSCM model, are given by
\begin{equation}\label{Ahat-SSCM}
\widehat{\Ab}=\widehat \Deltabf_{LS}^{-1/2}\widehat{\Vbf}_{(d)},
\end{equation}
\[
\widehat{\Bb}=\widehat{\Vbf}_{(d)}^{T}\widehat \Deltabf_{LS}^{1/2} \widehat\Sigmabf_{\Xbbb,\Fbbb}\widehat\Sigmabf_{\Fbbb,\Fbbb}^{-1},
\] 
\[
\widehat\mubf =  (\Xbb^T - \widehat\Ab \widehat\Bb \Fbb^T)\Hb^{-1}\ones_n (\ones_n^T\Hb^{-1}\ones_n)^{-1},
\]
and 
\begin{equation}\label{deltahat-SSCM}
\widehat \Deltabf =  \frac{1}{n}(\Xbbb^T -  \widehat \Ab \widehat \Bb \Fbbb^T) (\Xbbb^T - \widehat \Ab \widehat \Bb \Fbbb^T)^T,
\end{equation}
where $\widehat \Deltabf_{LS} = \frac{1}{n}(\Xbbb - \Fbbb \widehat{\Cb}_{LS}^T )^T (\Xbbb - \Fbbb \widehat{\Cb}_{LS}^T)$, $\widehat \Cb_{LS} = \frac{1}{n}\widehat\Sigmabf_{\Xbbb,\Fbbb}\widehat\Sigmabf_{\Fbbb,\Fbbb}^{-1}$ with $\widehat \Sigmabf_{\Xbbb,\Fbbb} = \frac{1}{n} \Xbbb^T \Fbbb$ and $\widehat\Sigmabf_{\Fbbb,\Fbbb}  = \frac{1}{n} \Fbbb^T \Fbbb$ and $\widehat{\Vbf}_{(d)}=[\widehat{\Vbf}_{1},\ldots,\widehat{\Vbf}_{d}]$ are the eigenvectors that corresponds to the j-th largest eigenvalues $\lambda_j^2$ of
$\widehat \Sigmabf \doteq  \widehat \Deltabf_{LS}^{-1/2}\widehat \Sigmabf_{\Xbbb,\Fbbb} \widehat\Sigmabf_{\Fbbb,\Fbbb}^{-1} \widehat\Sigmabf_{\Fbbb,\Xbbb}\widehat \Deltabf_{LS}^{-1/2}$ where
$\widehat \Sigmabf_{\Xbbb,\Fbbb}$ and  $\widehat\Sigmabf_{\Fbbb,\Fbbb}$ were given above and $ \widehat\Sigmabf_{\Fbbb,\Xbbb} = \widehat \Sigmabf_{\Xbbb,\Fbbb}^T$.
\end{prop}
The proof of this proposition is provided in Appendix \ref{AppendixA}.

For the correlation matrix $\Hbf$, a commonly assumed form is exponential (or spherical). Specifically, the elements of the correlation matrix $\Hbf_{ij}$ take the form $\Hbf_{ij} = \rho (d(\s_i - \s_j)) = \exp \big\{-\lambda (d(\s_i - \s_j)) \big \}$ for some parameter $\lambda$. To estimate this parameter, we maximize the likelihood in a grid of $\lambdabf = (\lambda_1, \ldots, \lambda_K)$ for some $K$. For each $\lambda_j$, $j=1,\ldots, K$ in the grid, we follow the steps described in this section to estimate the maximum likelihood estimators of the parameters $\mubf,\Ab, \Bb, \Deltabf$ and then  chose the $\lambda_j$ for which $l$ is maximized.

Finally, from Equation \eqref{reduction} and the estimators  $\widehat \Deltabf$ and $\widehat \Ab$ given in Equations \eqref{deltahat-SSCM} and \eqref{Ahat-SSCM}, respectively,  we obtain the estimated SDR for the SSCM model as
\begin{equation} \label{reduction-SSCM}
R(\Xbb) =  \Xbb \widehat \Deltabf^{-1} \widehat \Ab,
\end{equation}

\section{Spatial Autorregressive Error Model (SEM)}\label{secSEM}

Now to state the SEM model, the error term $\Ebb$ in model \eqref{spatial-model-1} follows an autoregressive structure of correlation; this is, 
\[
\Ebb = \theta \Wbf \Ebb + \Ubb,
\]
where $\theta$ is the coefficient of the spatially lagged error, $\Ubb \in {\real}^{n\times p}$ is such that $\Ubb^T \doteq (\ubf_{\s_1}, \ldots, \ubf_{\s_n})$ with $\ubf_{\s_i} \sim \mathcal N(\cerobf_p, \Deltabf), i= 1,\dots,n$ so that, $\vecc(\Ubb^T)  \sim \mathcal{N}(\cerobf_{np}, \ind_{n}\otimes\Deltabf)$. Here  $\cerobf_\ell$ is the $\ell$-dimensional vector of zeros. 

The matrix $\Wbf \in {\real}^{n\times n}$ is the {\it spatial weight matrix}, which quantifies the structure of spatial dependence, and $\theta$ is a spatial correlation parameter. There are several methods for computing $\Wbf$ (see \cite{Anselin88}). In this paper, we use the {\it normalized} weight matrix. First, we compute the maximum distance $d_{max}$ such that all points have at least one neighbor and then we use that distance to compute the neighborhoods. Then we define, for $i,j= 1,\dots,n$ $\Wbf_{ij} = 1$ if the distance between sites $i$ and $j$ is less or equal than $d_{max}$,, and $\Wbf_{ij} = 0$ otherwise. Finally, we normalize $\Wbf$ by columns such that the sum of the weights is 1.

Defining $\Wbf_{\theta} \doteq \ind_n - \theta \Wbf$ we have that,
\begin{equation*}
 \Ebb= (\ind_n - \theta \Wbf)^{-1}\Ubb \doteq \Wbf_{\theta}^{-1}\Ubb,
\end{equation*}
with 
\[
\vecc(\Ebb^T) = \vecc(\Ubb^T \Wbf_{\theta}^{-1}) = (\Wbf_{\theta}^{-1} \otimes \ind_n) \vecc(\Ubb^T).
\]
Now, since  $\vecc(\Ubb^T) \sim \mathcal{N}(\cerobf_{np}, \ind_{n}\otimes\Deltabf)$ results that
\begin{align*}
 \Cov(\vecc(\Ebb^T)) &= (\Wbf_{\theta}^{-1} \otimes \ind_n)  \Cov(\vecc(\Ubb^T)) (\Wbf_{\theta}^{-1} \otimes \ind_n) \\ &=  (\Wbf_{\theta}^{-1} \otimes \ind_n) (\ind_{n}\otimes\Deltabf) (\Wbf_{\theta}^{-1} \otimes \ind_n) \\ &= \Wbf_{\theta}^{-2} \otimes \Deltabf.
\end{align*}
Therefore, the SEM model is given by,
\begin{equation}\label{SEM-model-matrix}
\Xbb |\Y =  \ones_n  \mubf^T + \Fbb( \Ab \Bb)^T + \Ebb,
\end{equation}
with $\vecc(\Ebb^T) \sim \mathcal{N}(\cerobf_{np}, \Wbf_{\theta}^{-2} \otimes \Deltabf)$, which is model \eqref{spatial-model-1} with $\Sb = \Wbf_{\theta}^{-2}$. From this model we have that the log-likelihood function of the data is 
 
\begin{align} \label{log-likelihood-SEM}
	l(\Xbb|\Y;\mubf, \Ab, \Bb, \Deltabf, \theta)   &= - \frac{np}{2} \log (2\pi)  - \frac{n}{2} \log |\Deltabf| + p\log |\Wbf_{\theta}| \\
	&\hspace{-1cm} -\frac{1}{2} \tr \big(\Wbf_{\theta} (\Xbb -  \ones_n  \mubf^T - \Fbb\Bb^T \Ab^T) \Deltabf^{-1} (\Xbb -  \ones_n  \mubf^T - \Fbb\Bb^T \Ab^T)^T \Wbf_{\theta} \big). \nonumber
\end{align}
From this equation the maximum likelihood estimators for the parameters of the SEM model, and consequently the estimated SDR, can be obtained. This is stated in the following result. 
\begin{prop}\label{SEM-estimators}
Given the matrix $\Wbf$ and the coefficient $\theta$, the maximum likelihood estimators
 of the parameters of the SSCM model, are given by
\begin{equation}\label{Ahat-SEM}
\widehat{\Ab}=\widehat \Deltabf_{LS}^{-1/2}\widehat{\Vbf}_{(d)},
\end{equation}
\[
\widehat{\Bb}=\widehat{\Vbf}_{(d)}^{T}\widehat \Deltabf_{LS}^{1/2} \widehat\Sigmabf_{\Xbbb,\Fbbb}\widehat\Sigmabf_{\Fbbb,\Fbbb}^{-1},
\] 
\[
\widehat\mubf =  (\Xbb^T -\widehat \Ab \widehat\Bb \Fbb^T){\Wbf}_{\theta}^2\ones_n (\ones_n^T{\Wbf}_{\theta}^2\ones_n)^{-1},
\]
and 
\begin{equation}\label{deltahat-SEM}
\widehat \Deltabf =  \frac{1}{n}(\Xbbb^T -  \widehat \Ab \widehat \Bb \Fbbb^T) (\Xbbb^T - \widehat \Ab \widehat \Bb \Fbbb^T)^T,
\end{equation}
where $\widehat \Deltabf_{LS} = \frac{1}{n}(\Xbbb - \Fbbb \widehat{\Cb}_{LS}^T )^T (\Xbbb - \Fbbb \widehat{\Cb}_{LS}^T)$, $\widehat \Cb_{LS} = \frac{1}{n}\widehat\Sigmabf_{\Xbbb,\Fbbb}\widehat\Sigmabf_{\Fbbb,\Fbbb}^{-1}$ with $\widehat \Sigmabf_{\Xbbb,\Fbbb} = \frac{1}{n} \Xbbb^T \Fbbb$ and $\widehat\Sigmabf_{\Fbbb,\Fbbb}  = \frac{1}{n} \Fbbb^T \Fbbb$ and $\widehat{\Vbf}_{(d)}=[\widehat{\Vbf}_{1},\ldots,\widehat{\Vbf}_{d}]$ are the eigenvectors that corresponds to the j-th largest eigenvalues $\lambda_j^2$ of
$\widehat \Sigmabf \doteq  \widehat \Deltabf_{LS}^{-1/2}\widehat \Sigmabf_{\Xbbb,\Fbbb} \widehat\Sigmabf_{\Fbbb,\Fbbb}^{-1} \widehat\Sigmabf_{\Fbbb,\Xbbb}\widehat \Deltabf_{LS}^{-1/2}$ where
$\widehat \Sigmabf_{\Xbbb,\Fbbb}$ and  $\widehat\Sigmabf_{\Fbbb,\Fbbb}$ were given above and $ \widehat\Sigmabf_{\Fbbb,\Xbbb} = \widehat \Sigmabf_{\Xbbb,\Fbbb}^T$.
\end{prop}
The proof of this proposition is presented in Appendix \ref{AppendixB}.

For the matrix ${\Wbf}_{\theta}$, we need to estimate the parameter $\theta$. To do this, it is common to  maximize the likelihood over  a grid of $\thetabf = (\theta_1, \ldots, \theta_K)$ for some $K$. For each $\theta_j$, $j=1,\ldots, K$ in the grid, we follow the steps described in this section to estimate the maximum likelihood estimators of the parameters $\mubf,\Ab, \Bb, \Deltabf$ and then we chose the $\theta_j$ for which $l$ is maximized.

Finally, using the estimators $\widehat \Deltabf$ and $\widehat \Ab$ as given in Equations \eqref{deltahat-SEM} and \eqref{Ahat-SEM}, respectively and from the  Equation \eqref{reduction}, the estimated SDR for the SEM model is given by 
\begin{equation} \label{reduction-SEM}
R(\Xbb) =  \Xbb \widehat \Deltabf^{-1} \widehat \Ab,
\end{equation}

 
\section{Spatial prediction}\label{prediction}

To predict the response $Y_{\s} $ in a point $\s_0$  given a covariates vector  $\x_{\s_0}$, we will use  nonparametric estimation of the spatial regression model \eqref{genmodel-2}, that is  $\E_{Y_{\s} | \X_{\s}}(Y_{\s} | \X_{\s})=\widetilde{\eta}(R(\X_{\s}))$ for a sufficient regression $R(\X_{\s})$ of $Y_{\s}$ given $\X_{\s}$. For this purpose, given the sample $\{(\X_{\s_i}, Y_{\s_i})\}_{i=1}^n$, we predict the response variable $Y$ in a unsampled location $\s_0$ as
\begin{equation*}\label{estimator2}
\widehat {Y}_{\s_0} = \widehat \E_{Y_{\s_{0}} | \X_{\s}}(Y_{\s_0} | \X_{\s_0} = \x_{\s_0}) = \sum_{i=1}^n w_{i,n}(R(\x_{\s_0})) Y_{\s_i},
\end{equation*}
where  $w_{i,n}(R(\x_{\s_0}))$ are probability weights valued at $R(\x_{\s_0})$. For these weights we will consider two alternatives, the first one is the classical Nadaraya-Watson kernel estimator
\begin{equation}\label{weights4}
w_{i,n}(\x_{\s_0}) = \frac{K_1\left(\frac{||\widehat R(\x_{\s_0}) - \widehat R(\X_ {s_i})||}{h_1}\right)}{\sum_{j=1}^n K_1\left(\frac{||\widehat R(\x_{\s_0}) - \widehat R(\X_ {s_j})||}{h_1}\right)},
\end{equation}
where $K_1$ is a \textit{Kernel} function and $h_1$ is the smoothing parameter. And, in the second alternative (see \cite{DN16}) we add to the weights (\ref{weights5}) a second kernel $K_2$ which takes into account the spatial locations of the data, 
\begin{equation}\label{weights5}
w_{i,n}(\x_{\s_0}) = \frac{ K_1\left(\frac{||\widehat R(\x_{\s_0}) - \widehat R(\X_ {s_i})||}{h_1}\right) K_2\left(\frac{||s_0-s_i||}{h_2}\right)}{\sum_{j=1}^n K_1\left(\frac{||\widehat R(\x_{\s_0}) - \widehat R(\X_ {s_j})||}{h_1}\right) K_2\left(\frac{||s_0-s_j||}{h_2}\right)},
\end{equation}
where, as before, $K_2$ is a \textit{Kernel} function and $h_2$ is the smoothing parameter. 

In all cases, the kernels  $K_1$ and $K_2$ are chosen as Gaussian kernels, and their respective smoothing parameters are determined via leave-one-out cross-validation on the training sample.

The methods will be named based on how we estimate the reduction and the weights we use. For the single kernel weights \eqref{weights4}, we will use the prefix \texttt{1k}, and for the two-kernel weights \eqref{weights5}, the prefix \texttt{2k}. Accordingly, if the reduction is estimated using the new methods described in this paper, the methods will be labeled as \texttt{1k.SSCM}, \texttt{1k.SEM}, \texttt{2k.SSCM}, and \texttt{2k.SEM}. If we estimate the reduction under the assumption of data independence, following the Principal Fitted Component (PFC) framework of \cite{CookForzani2008}, the notation \texttt{1k.Ind} and \texttt{2k.Ind} will be used. 

Finally, all these methods will be compared with the classical nonparametric  prediction method computed with the full set of covariates (this is, without reduction), given by 
\begin{equation}\label{weights5}
w_{i,n}(\x_{\s_0}) = \frac{K_1\left(\frac{||\x_{\s_0} - \X_ {s_i}||}{h_1}\right)}{\sum_{j=1}^n K_1\left(\frac{||\x_{\s_0} - \X_ {s_j}||}{h_1}\right)},
\end{equation}
for one kernel estimator and
\begin{equation*}
w_{i,n}(\x_{\s_0}) = \frac{K_1\left(\frac{||\x_{\s_0} - \X_ {s_i}||}{h_1}\right) K_2\left(\frac{||s_0-s_i||}{h_2}\right)}{\sum_{j=1}^n K_1\left(\frac{||\x_{\s_0} - \X_ {s_j}||}{h_1}\right) K_2\left(\frac{||s_0-s_j||}{h_2}\right)},
\end{equation*}
for the two kernel predictor. This two alternatives will be call \texttt{1k.FULL}  and \texttt{2k.FULL}, respectively. 

\section{Choosing the dimension $d$}

In the models presented in Sections \ref{secSSCM} and \ref{secSEM}, we assume that the dimension  $d$ of the reduction is known. However, in practice, $d$ must be selected based on some criteria. Given our model-based SDR approach and the maximum likelihood framework, we can apply likelihood-based criteria as in \cite{CookForzani2008}, where the choice of the dimension will be
independent of the regression function used for prediction. In particular, we can use the likelihood ratio (LR) statistics, $ \Lambda_{\delta} = 2(L_{\text{min}(r,p)} - L_{\delta})$, to test the null hypothesis $d = \delta$ against the alternative $d > \delta$. Under the null hypothesis, $ \Lambda_{\delta}$ is asymptotically distributed as a chi-square  $\chi^2_q$, with $q = (r - \delta)(p - \delta)$. Starting from $\delta = 0$, the likelihood ratio test is performed sequentially, selecting $d$ when the null hypothesis is no longer rejected. Alternatively, information criteria such as AIC or BIC can be used to select the optimal $d$ by minimizing either
\[
AIC(\delta) = -2L_{\delta} + 2\left( \frac{p(p+3)}{2} + r\delta + \delta(p-\delta) \right)
\]
or
\[
BIC(\delta) = -2L_{\delta} + \log(n)\left( \frac{p(p+3)}{2} + r\delta + \delta(p-\delta) \right),
\]
for $\delta = 0, \ldots, \min(r,p)$.

On the other hand, given that our primary goal is prediction, an alternative approach could be to select the dimension that minimizes the cross-validated prediction error across a range of $d$ values \citep{forzani18}. This criteria will be denoted CV-MPE (Minimum Prediction Error). 
In this case, unlike likelihood criteria, the selection of $d$ will depend on the predictive rule.

\section{Simulation studies}\label{simulations}

To demonstrate the performance of the prediction methods, this section presents several simulation studies where we vary the sample size ($n$), the number of covariates ($p$), and the hyperparameters $r$ and $d$. 

\subsection{Settings}

First, we randomly generate sample point locations $\s = (s_1, s_2) \in \mathbb{R}^2$ within the square $[0,1] \times [0,1]$. This procedure results in an irregular grid that serves as the basis for simulating our data sets. Once the grid is established, we simulate the spatial model described in Equation (\ref{model-last}): 
\[
\X|Y= \mubf + \Ab\Bb \f_{Y}+\epsilonbf,
\]
where $\mubf\in \real^p$ is a vector of $p$ standard normal numbers; $\Ab\in\real^{p\times d}$ and $\Bb\in\real^{d\times r}$ are full-rank  matrices generated with standard normal numbers; $\Deltabf$ is also generated with standard normal numbers and transformed to be symmetric and positive-definite; $\f_{Y_{\s}} \in \real^{r\times 1}$ is a polynomial  of degree $r$ of $Y_{\s}$, where $Y_{\s}$ is generated using a Gaussian random field with trend, with $E(Y_{\s})= 1+0.1s_1+0.05s_2$ and spherical covariance function characterized by a sill of $1.25$ and a range of $2$.

For $\X$, we considered two scenarios: $\X$ generated under SSCM model with $\lambda=0.1$ and $\X$ generated under SEM model with  $\theta=0.8$. For the SEM model, as described earlier, $\W$ is the normalized weight matrix, where $\Wbf_{ij} = 1$ if the distance between sites $i$ and $j$ is less than or equal to the maximum distance $d_{max}$, ensuring that each point has at least one neighbor. Otherwise, $\Wbf_{ij} = 0$. The matrix $\Wbf$ is then column-normalized so that the sum of weights in each column equals 1.

This procedure is repeated $100$ times, with each iteration  involving a split of the data into two portions: $70\%$ of the data is used for parameter estimation (training sample), while the remaining $30\%$ is used to compute the Mean Square Error (MSE) on the test sample.  At the end of the simulation, we obtain $100$ MSE value.

\subsection{Results}
\subsubsection{Dimension reduction}
In order to show the advantages of dimension reduction and compare the methodologies designed for spatial data, first we keep fixed the parameters $p = 24$, $n_1 = n_2 = 20$ so that $n = 20 \times 20 = 400$,  and vary the parameters $r = 2,3$ and $d =1,2$. Table \ref{difdyr} shows the average MSE computed over the 100 replications for each prediction method under SSCM and SEM models. The minimum errors for each combination of  $r$ and $d$ are highlighted in bold. At first, it can be observed that, in general, the dimension reduction yields better results compared to using the full set of predictors without reduction. Additionally, reduction methods that include spatial dependencies perform better compared to the PFC method under the independence assumption (i.e. \texttt{1k.Ind} and \texttt{2k.Ind}), except for the combination of $d=1$ and $r=3$ under SSCM, where \texttt{2k.Ind} achieves the lowest MSE, though it is very comparable to  \texttt{2k.SSCM}  and \texttt{2k.SEM}. 

For data simulated under SSCM model, \texttt{2k.Ind}, \texttt{2k.SSCM} and \texttt{2k.SEM} yield the same lowest MSE when $d=2$ and $r=2$. When $d=2$ and $r=3$ the  best predictive results are provided by  the \texttt{2k.SSCM}. Additionally, for $d=1$ and $r=2$, both \texttt{2k.SEM} and \texttt{2k.SSSC} achive identical MSE, resulting in the best predictive performance. Under SEM model, when $d=2$ the best predictive results are obtained with dimension reduction method \texttt{2k.SSCM}, whereas for $d=1$ better results are obtained with  \texttt{2k.SEM}. Therefore, these simulation results show that generating spatial data under SEM or SSCM does not always guarantee better predictive performance when assuming these respective models in inverse model estimates. However, even if a spatial model is not the correct one, assuming it generally tends to result in better predictions compared with an independence model.

Another remarkable result to observe in this simulation is the advantages of using two kernel predictors for all dimension reduction methodologies as well as in regression without reduction (\texttt{FULL}). In this direction and for the easy of exposition, in the following simulations we will show just the two kernels results. 

\begin{table}
\begin{center}
\begin{tabular}{c|cc|cc|cc|cc}
\hline	
&\multicolumn{4}{c|}{Under SSCM} & \multicolumn{4}{c}{Under SEM}\\\hline
&\multicolumn{2}{c|}{$r=2$} & \multicolumn{2}{c|}{$r=3$} &  \multicolumn{2}{c|}{$r=2$} & \multicolumn{2}{c}{$r=3$} \\\hline
&        $d=1$ &       $ d=2$ &        $d=1$ &        $d=2$ &        $d=1$ &       $ d=2$ &        $d=1$ &        $d=2$ \\\hline \hline
\multirow{2}{*}{\texttt{1k.FULL}}&0.041&0.018&0.032&0.009&0.173&0.109&0.160&0.131\\
&(0.067)&(0.036)&(0.044)&(0.011)&(0.151)&(0.126)&(0.160)&(0.183)\\ \hline
\multirow{2}{*}{\texttt{2k.FULL}}&0.017&{ 0.012}&0.015&0.007&0.041&0.036&0.036&0.033\\
&(0.014)&(0.020)&(0.014)&(0.007)&(0.013)&(0.017)&(0.016)&(0.013)\\ \hline
\multirow{2}{*}{\texttt{1k.Ind}}&0.103&0.034&0.074&0.014&0.091&0.021&0.093&0.038\\
&(0.221)&(0.073)&(0.126)&(0.033)&(0.133)&(0.047)&(0.143)&(0.061)\\ \hline
\multirow{2}{*}{\texttt{2k.Ind}}&0.015&{\bf 0.008}&{\bf 0.013}&0.005&0.015&0.008&0.014&0.010\\
&(0.029)&(0.023)&(0.017)&(0.007)&(0.013)&(0.017)&(0.013)&(0.012)\\ \hline
\multirow{2}{*}{\texttt{1k.SSCM}}&0.088&0.029&0.082&0.012&0.203&0.026&0.213&0.035\\
&(0.159)&(0.070)&(0.142)&(0.068)&(0.187)&(0.048)&(0.214)&(0.062)\\ \hline
\multirow{2}{*}{\texttt{2k.SSCM}}&{\bf 0.014}&{\bf 0.008}&0.015&{\bf 0.004}&0.030&{\bf 0.007}&0.031&{\bf 0.009}\\
&(0.031)&(0.021)&(0.023)&(0.006)&(0.017)&(0.007)&(0.018)&(0.011)\\ \hline
\multirow{2}{*}{\texttt{1k.SEM}}&0.091&0.029&0.082&0.017&0.083&0.019&0.078&0.033\\
&(0.171)&(0.068)&(0.143)&(0.037)&(0.133)&(0.046)&(0.132)&(0.056)\\ \hline
\multirow{2}{*}{\texttt{2k.SEM}}&{\bf 0.014} &{\bf 0.008} &0.014&0.007&{\bf 0.011} &{\bf 0.007}&{\bf 0.011} &0.010\\ 
&(0.031)&(0.023)&(0.023)&(0.010)&(0.011)&(0.015)&(0.010)&(0.012)\\ \hline
\end{tabular}  
\end{center}
\caption{Mean and standard deviation of the cross-validated MSE (over $100$ replications) under SSCM and SEM models with $n=400$ and $p=24$, computed over $100$ replications of any prediction method for different values of the dimension $d$ and also $r$. Standar deviations are reported in parenthesis.}\label{difdyr}
\end{table}

\subsubsection{Effects of Sample size}

\begin{figure}[] 
\begin{multicols}{2}
    \includegraphics[width=0.8\linewidth]{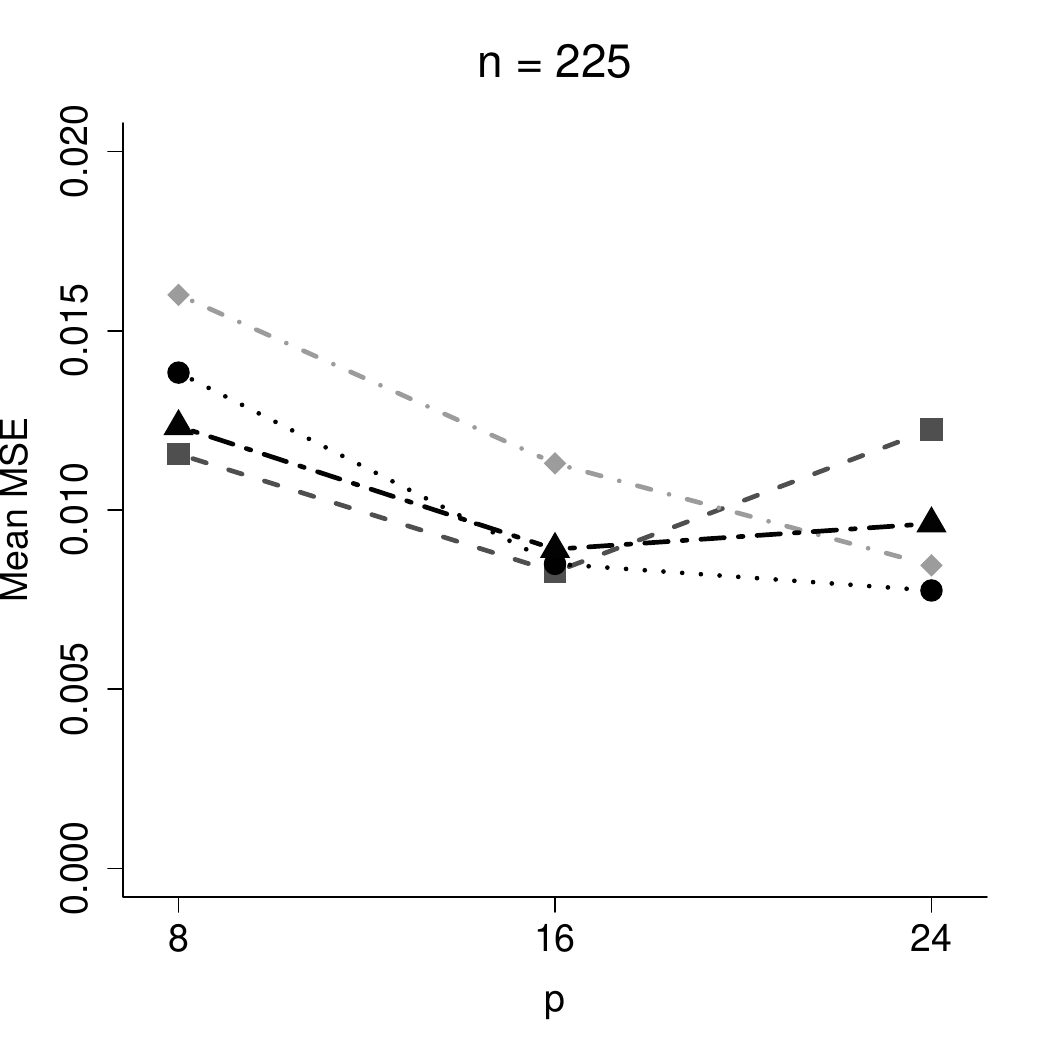}\par 
    \includegraphics[width=0.8\linewidth]{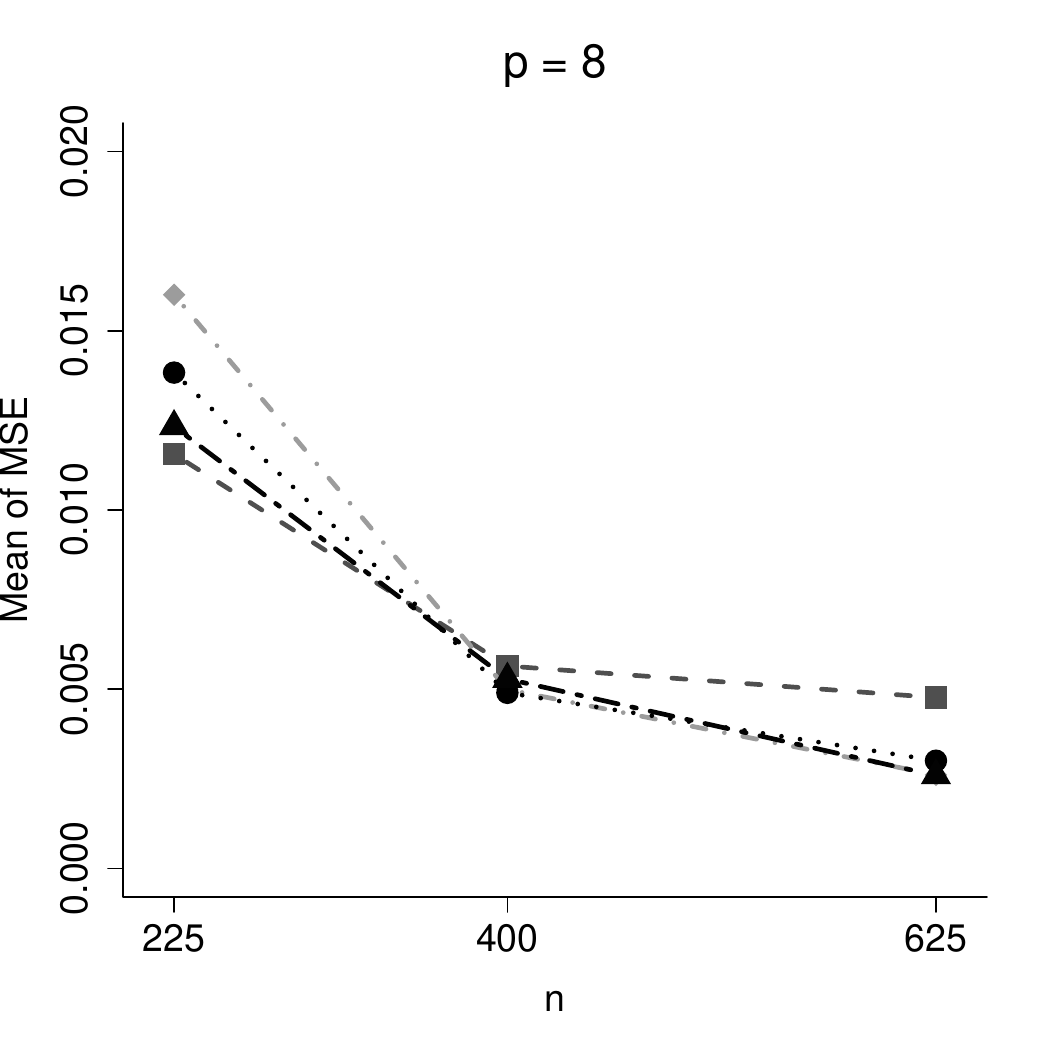}\par 
    \end{multicols}
\begin{multicols}{2}
    \includegraphics[width=0.8\linewidth]{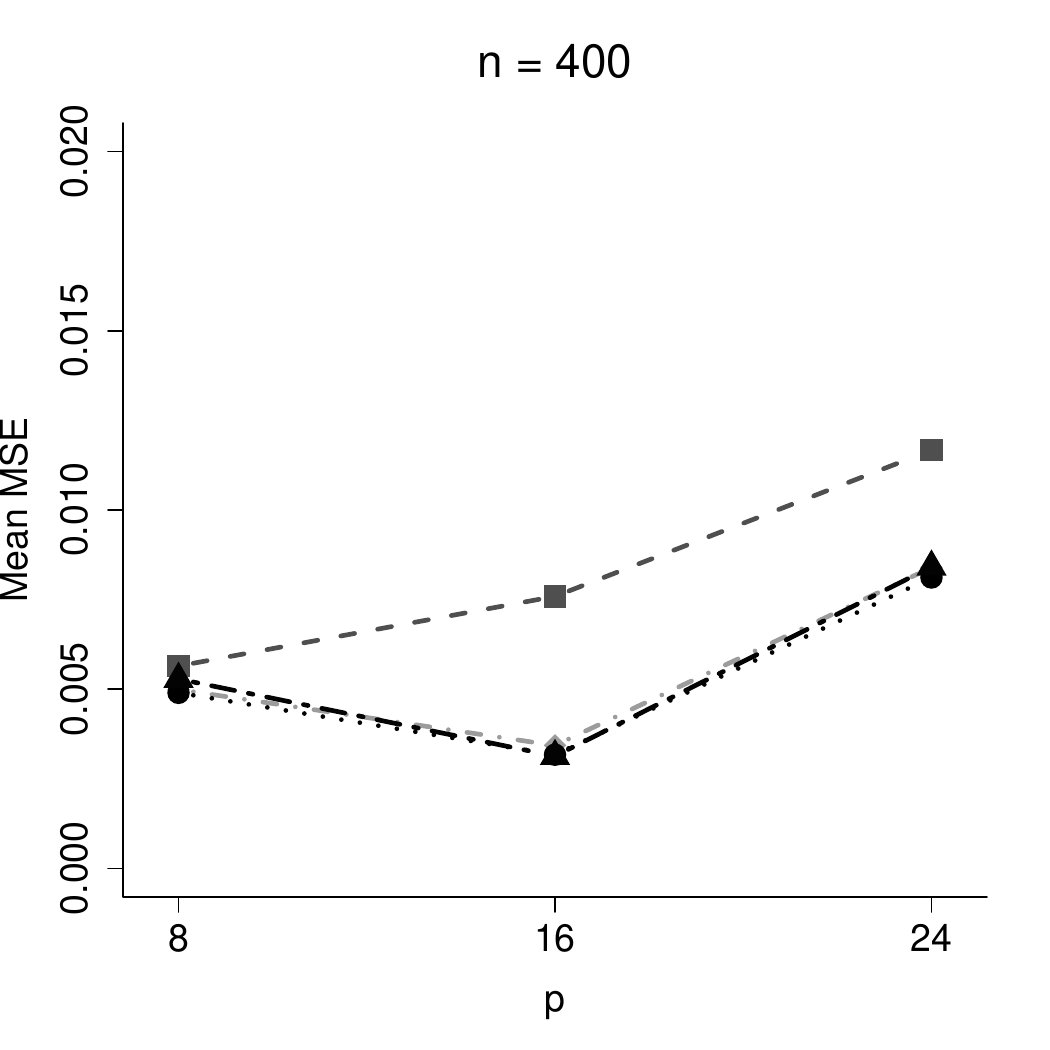}\par
    \includegraphics[width=0.8\linewidth]{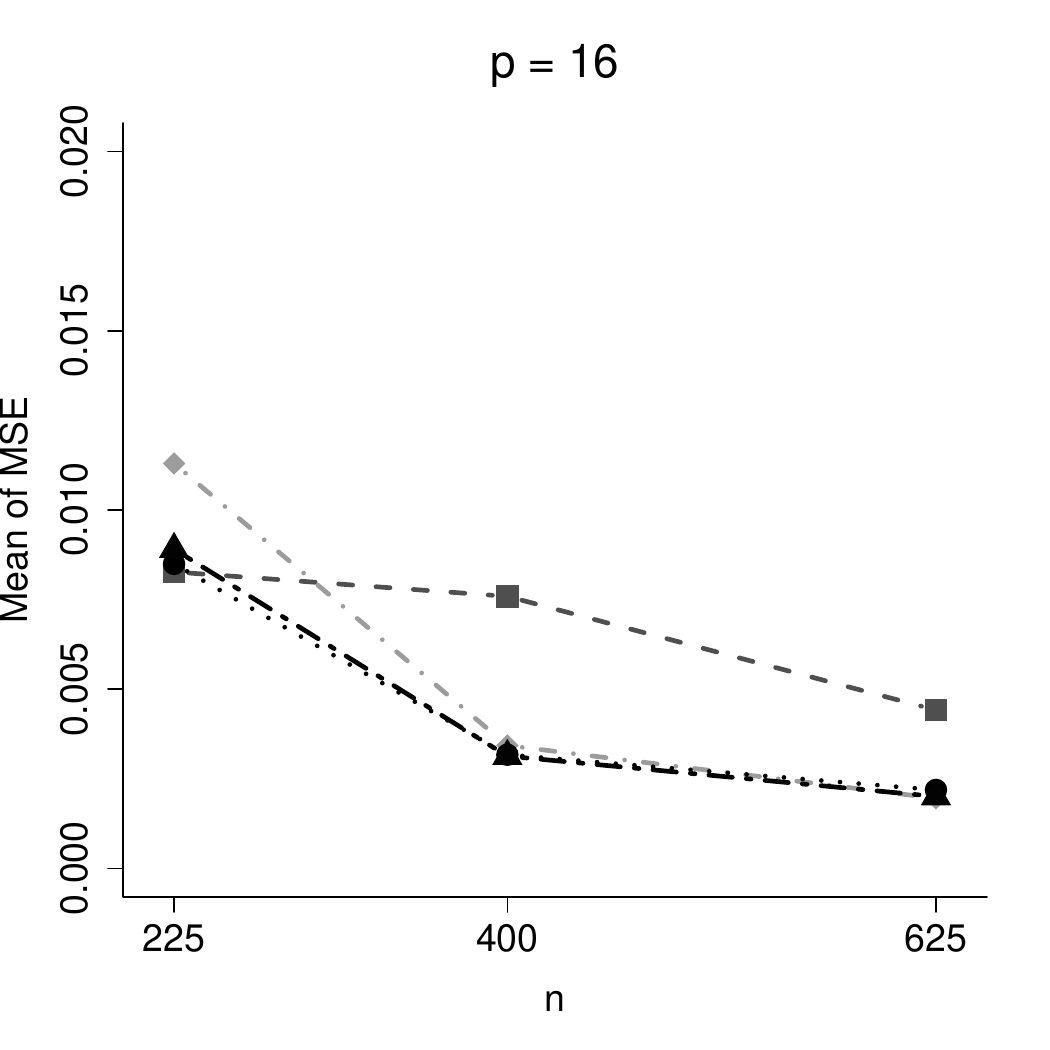}\par
\end{multicols}
\begin{multicols}{2}
    \includegraphics[width=0.8\linewidth]{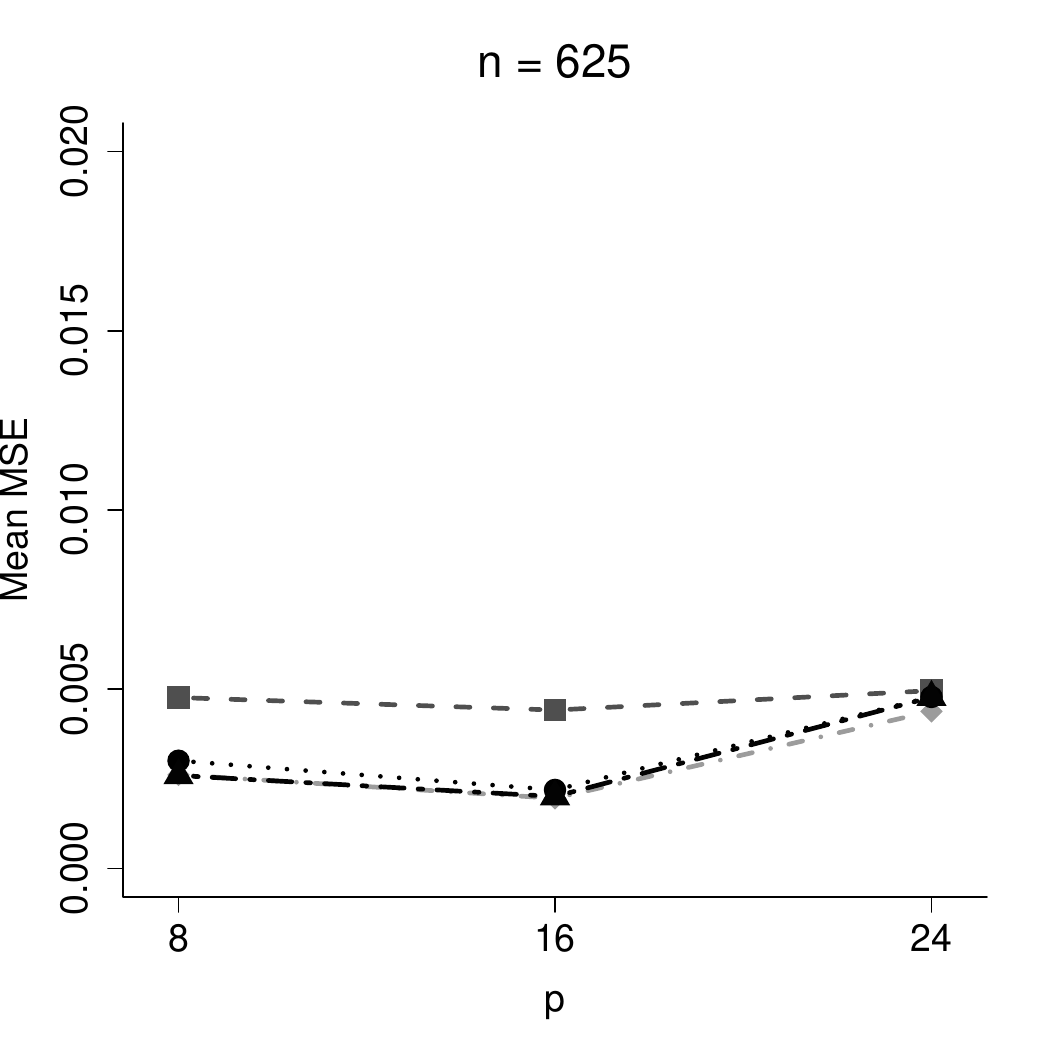}\par
    \includegraphics[width=0.8\linewidth]{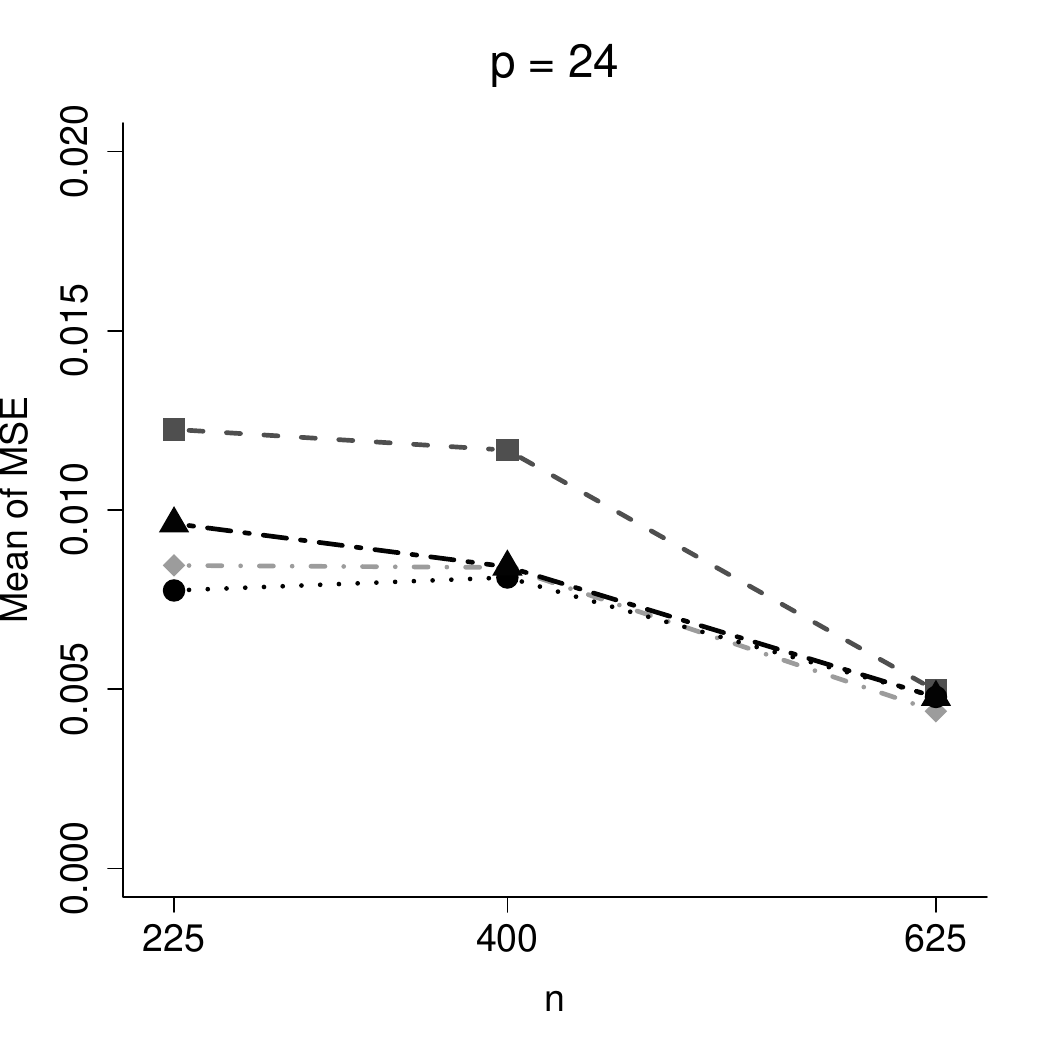}\par
\end{multicols}
\hspace{3cm}\includegraphics[width=0.7\linewidth]{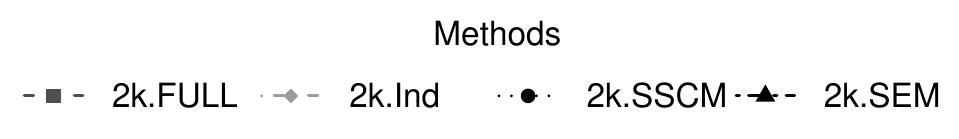}
\caption{Average cross-validated MSE (computed over 100 replications) for the SSCM model using different prediction methods across various sample sizes.}
\label{plot-SSCM}
\end{figure}

\begin{figure}[] 
\begin{multicols}{2}
    \includegraphics[width=0.8\linewidth]{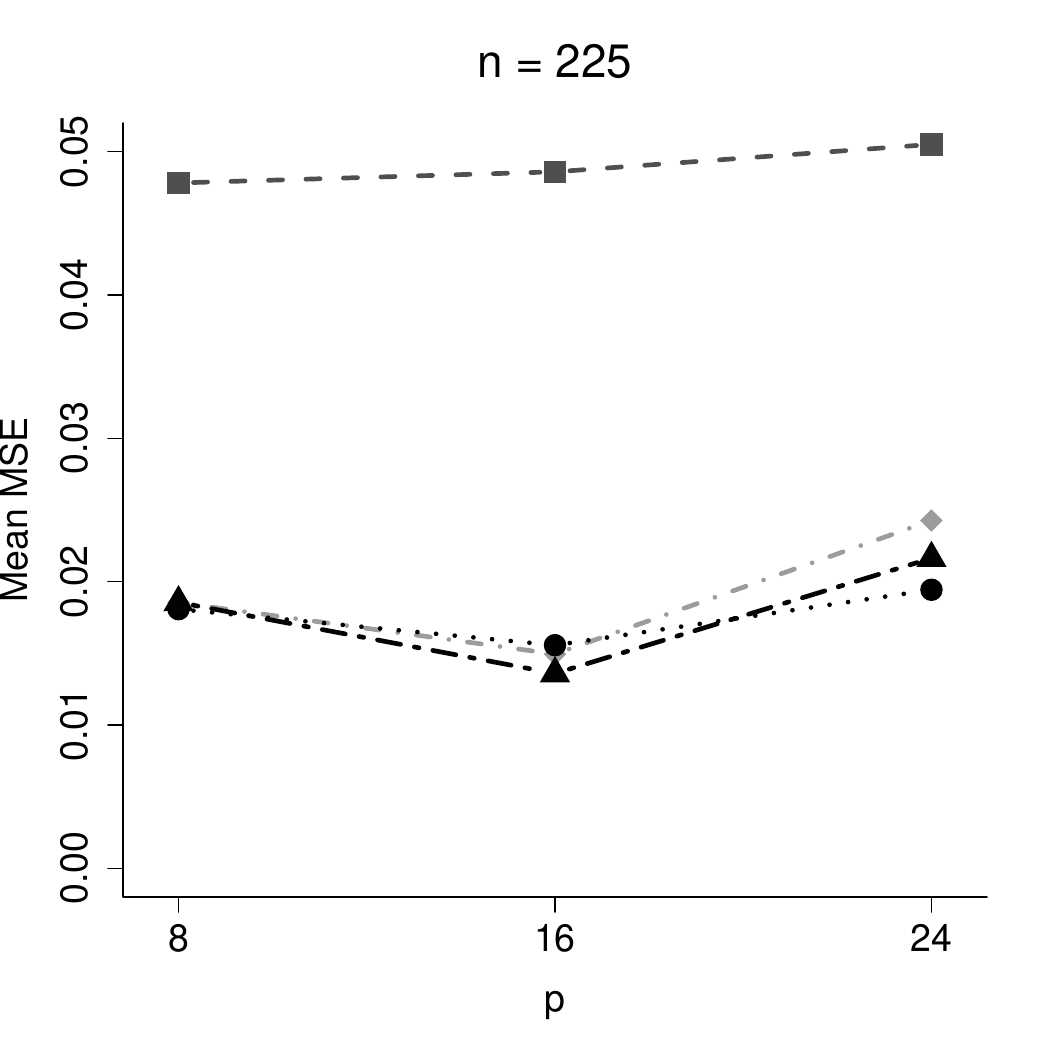}\par 
    \includegraphics[width=0.8\linewidth]{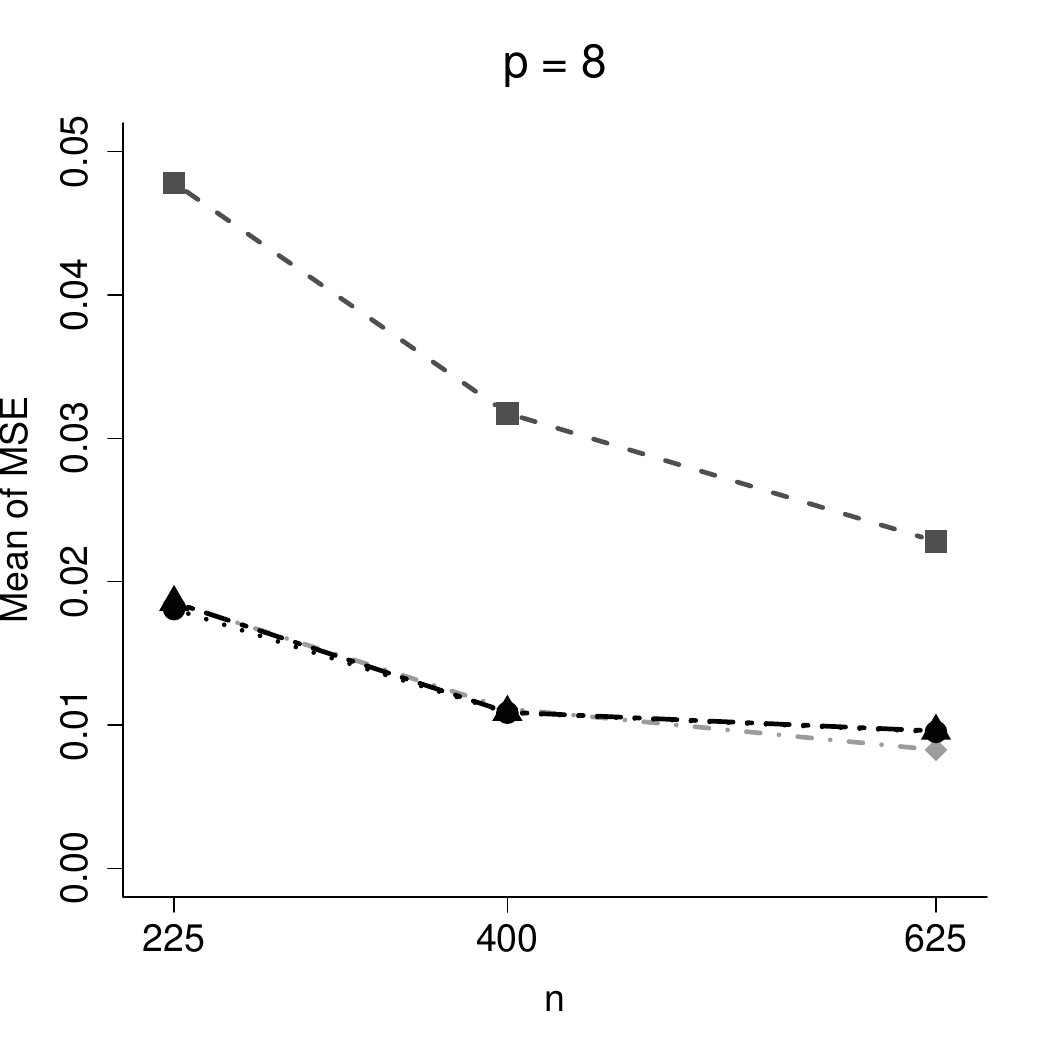}\par 
    \end{multicols}
\begin{multicols}{2}
    \includegraphics[width=0.8\linewidth]{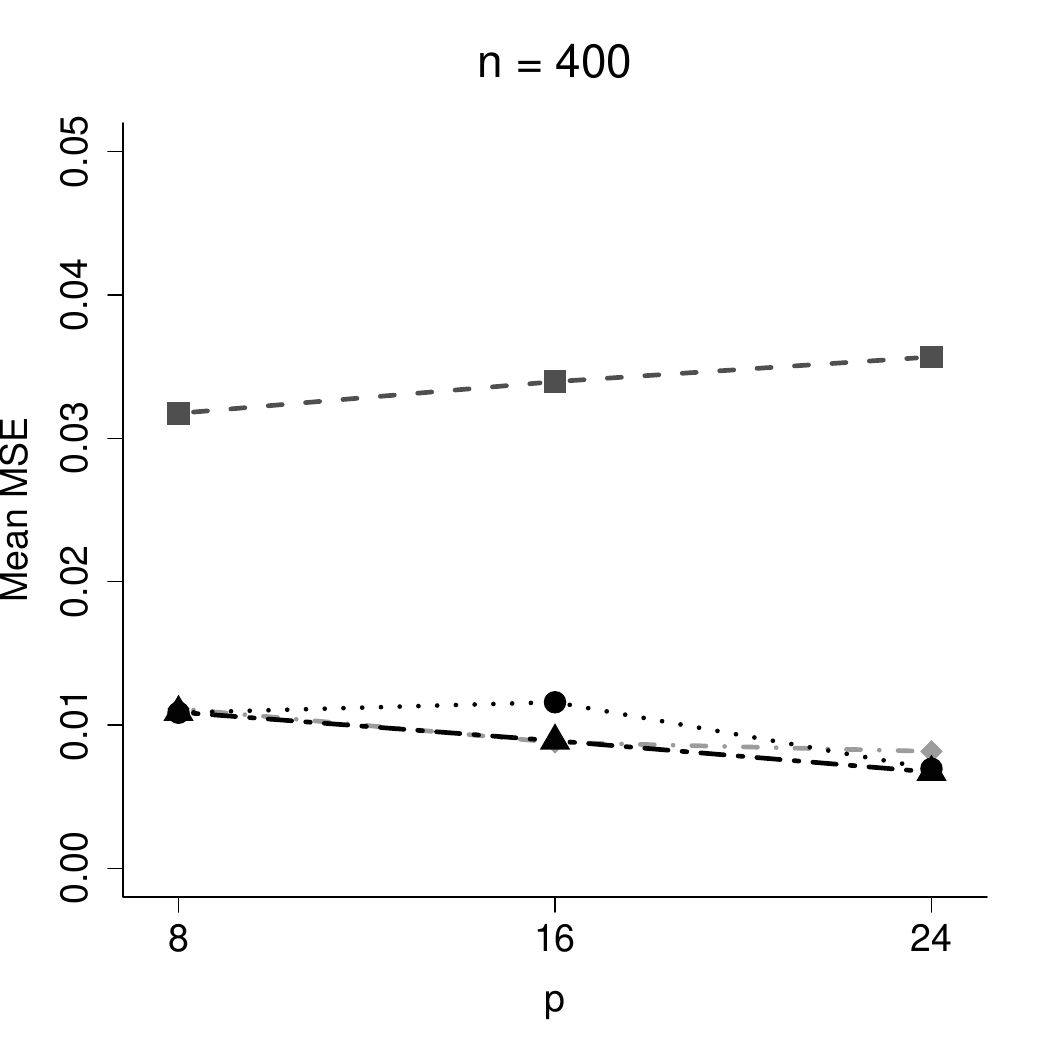}\par
    \includegraphics[width=0.8\linewidth]{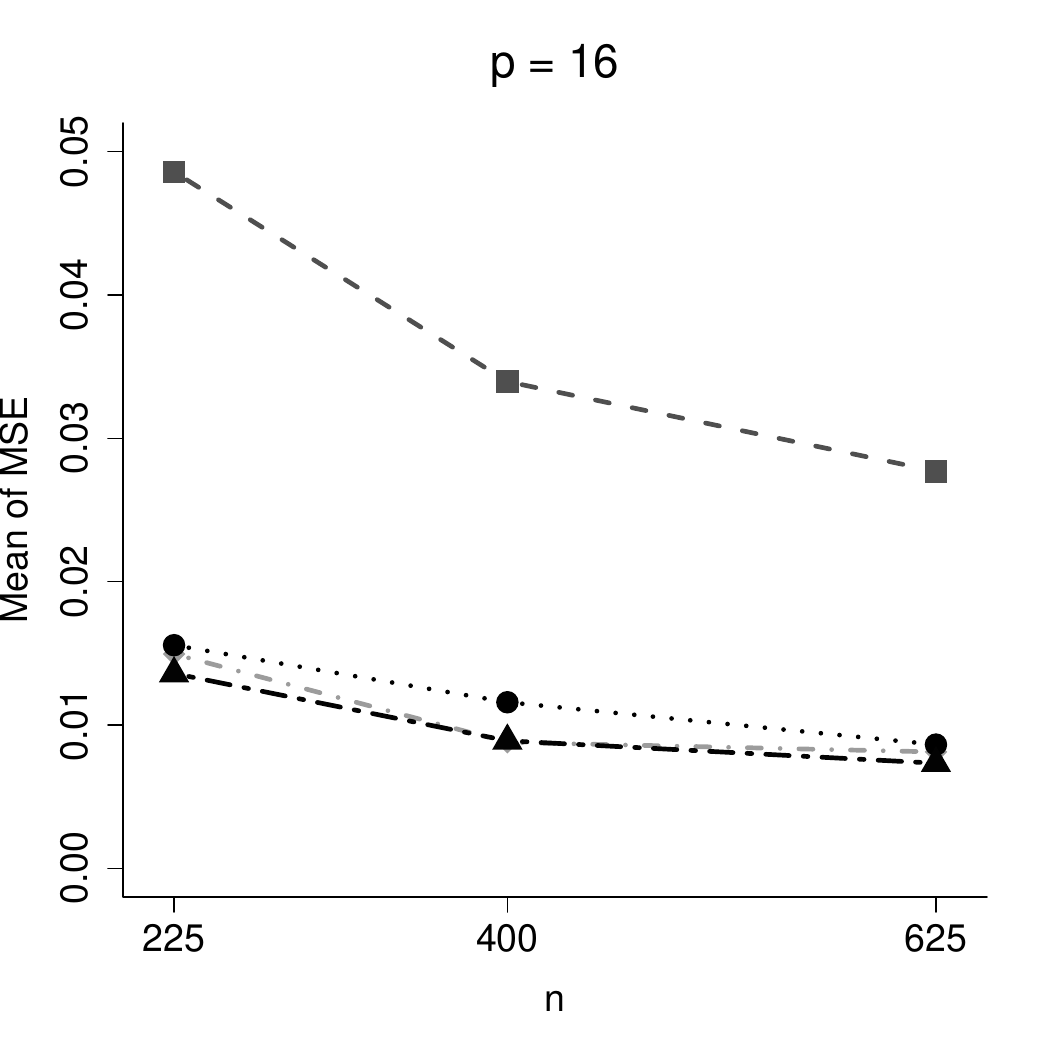}\par
\end{multicols}
\begin{multicols}{2}
    \includegraphics[width=0.8\linewidth]{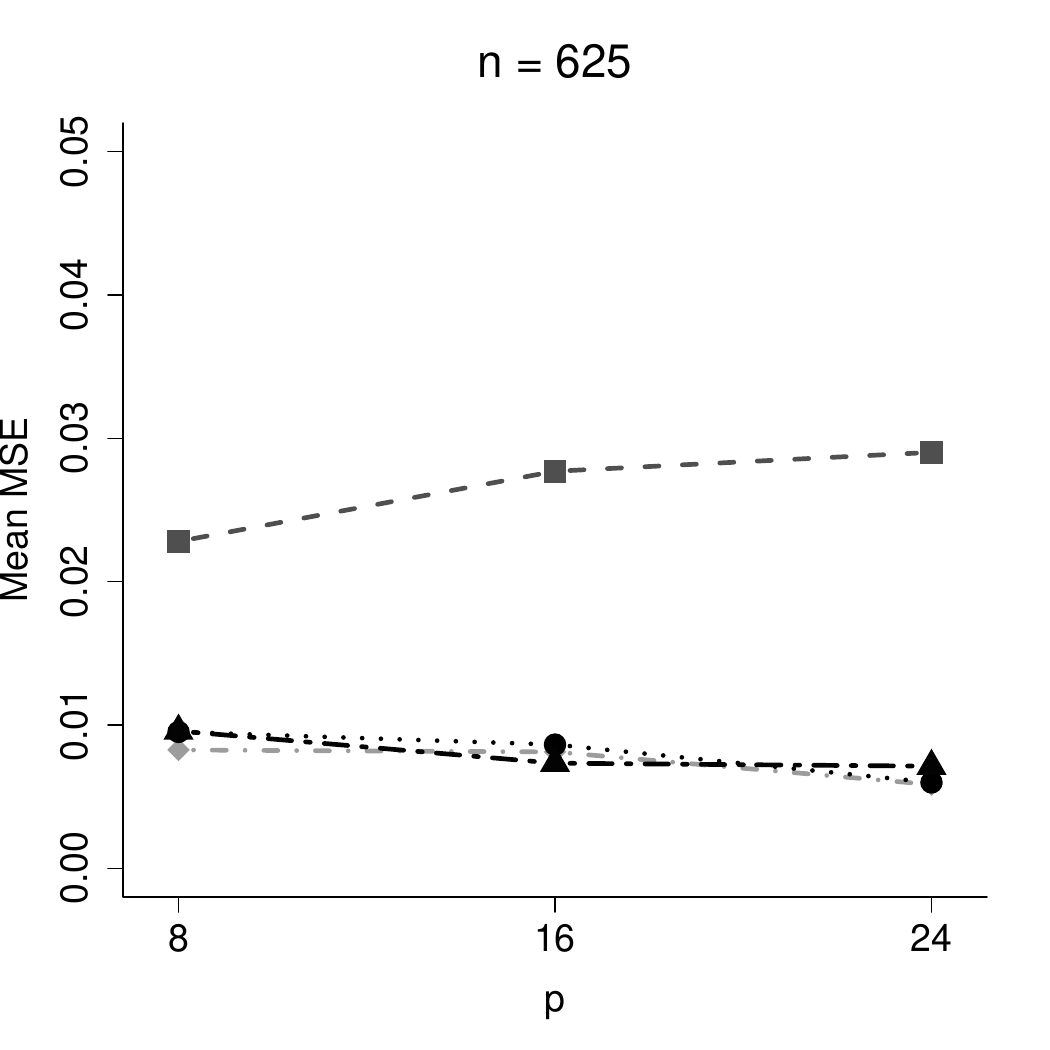}\par
    \includegraphics[width=0.8\linewidth]{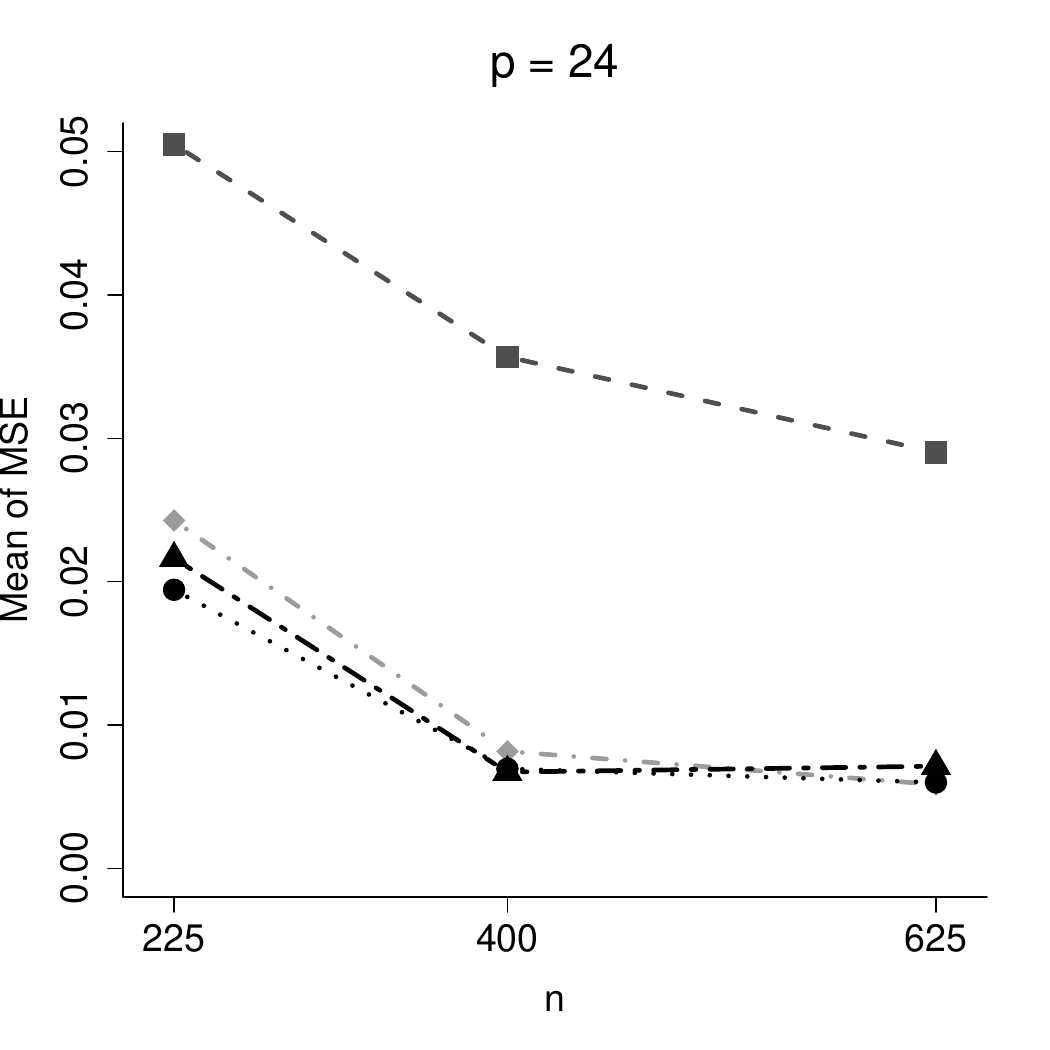}\par
\end{multicols}
\hspace{3cm}\includegraphics[width=0.7\linewidth]{legenda_horizontal.pdf}
\caption{Average cross-validated MSE (computed over 100 replications) for the SEM model using different prediction methods across various sample sizes.}
\label{plot-SEM}
\end{figure}

In this case, we study how predictive errors behave when the sample size $n$ changes (i.e., $n_1$ and $n_2$) and for different numbers of covariates, that is, $p = 8,16,24$. For the easy of exposition, we set $r = 2,d = 2$ for all cases. Right columns in Figures \ref{plot-SSCM} for SSCM and \ref{plot-SEM} for SEM,
show that, for each $p$, the mean of the MSE computed over $100$ replications decreases when the sample size increases. Under the SSCM model,  when $n=225$ it can be observed that, for $p=8$ and $p=16$, \texttt{2k.FULL} performs better than dimension reduction methodologies. However, as the sample size increases, dimension reduction methodologies quickly outperform predictions using the full set of predictors. Additionally, the improvements from considering sufficient dimension reduction with spatial modeling become more noticeable with smaller sample sizes. Under the SEM model, the advantages of using dimension reduction techniques are more evident, and the performance gap between them remains considerable for all sample sizes. In this case, the gains in prediction using spatial dimension reduction are greater for $p=24$, with very comparable errors to the independent PFC method as the sample size increases.

In summary, as the sample size increases, the prediction errors generally tend to decrease. The spatial reduction methodologies exhibit very comparable errors among themselves and with respect to the reduction method under the independence assumption. However, for the smaller sample size, spatial reduction methods yield lower predictive errors.  In all cases, they outperform predictions without reduction.

\subsubsection{Effects of number of covariates}

In the left columns of Figures \ref{plot-SSCM} and \ref{plot-SEM}, can be observe how the prediction errors change when the number of covariates increases for different sample sizes. Again, we set $r = 2,d = 2$ for both, SSCM and SEM models.\\
Under SSCM, for $n=225$, it can be seen that non-reduction methods has perform better than reduction methodologies when $p=8$. However, as the number of covariates increases  to $p=24$, the MSE for reduction methodologies decreases, indicating better predictive results for these methods, particularly with the \texttt{2k.SSCM}. When considering higher sample sizes, different behavior is observed.  The MSE grows noticeably for $n=400$ as the number of covariates increases, especially for the non-reduction approach, which deteriorates more quickly than predictors with dimension reduction techniques. For $n=625$, a more stable behavior is observed, where dimension reduction methodologies consistently outperform non-reduction methods, although the MSE values converge as $p$ reaches $24$.\\
Simulations under the SEM model show a considerable higher MSE for non-reduction predictor respect to dimension reduction methodologies, with an increasing trend as $p$ increases. Except for the case of $n=225$, reduction methodologies consistently have lower prediction errors as $p$ grows.

\section{Real Data Applications}\label{realdata}
To evaluate our model-based spatial dimension reduction methods using the two nonparametric spatial predictors, we present three examples from different fields of application with different spatial characteristics. These examples allow us to show the behaviour of our methods for different combinations of sample size ($n$), number of predictors ($p$), and for different domains (physical/geological, socio-economic) as well as observational scales, whether points or areas (with coordinate points representing polygon centroids).

Specifically, in the first example, we use the classical Meuse data set to study the zinc concentration prediction problem with five predictor variables. In the second example, we study dimension reduction methods for a composite index construction for predictive purposes. Here we take 25 variables to construct a school characteristics index to predict the average of 4th grade scores based on school districts in Ohio. In the last example, taking a data set on 72 countries, we analyse the performance of our spatial models for dimension reduction methods to predict GDP growth rates over the 1960–80 period using 19 predictors about the macroeconomic, demographic, social, cultural, and institutional characteristics of these countries.

As in the simulations, to ensure robust results the process of dividing the whole data set in training and testing samples is repeated 100 times. As before, $70\%$ of the data is used as the training sample for parameter estimation and dimension, while the remaining $30\%$ is reserved as the testing sample, computing the MSE.

\subsection{Predicting Zinc Concentration: The Meuse Data Set}
In geostatistics, a very popular data set used to evaluate interpolation and prediction methods is the Meuse data set provided by \texttt{sp} and \texttt{gstat} [R] packages. From this data set, the goal is to predict the \textit{zinc concentration} ($Y$)  in the top soil in a flood plain along the river Meuse. In addition to locations and zinc concentration measures, we have others soil and landscape variables that we use as predictors: topsoil cadmium concentration ($X_1$), topsoil copper concentration ($X_2$), topsoil lead concentration ($X_3$), relative elevation above local river bed ($X_4$) and distance to river Meuse in metres ($X_5$). The point data set contains 155 samples of zinc concentration with observations on predictor variables. As is common when analyzing this dataset, we apply a logarithmic transformation to the response variable to improve its behavior. Therefore, our target variable to predict is $Y\equiv \log(\text{zinc}_{\mathbf{s}_0})$ in a location point $\mathbf{s}_0$.
\begin{figure}[]
	\centering
	\includegraphics[scale=0.75]{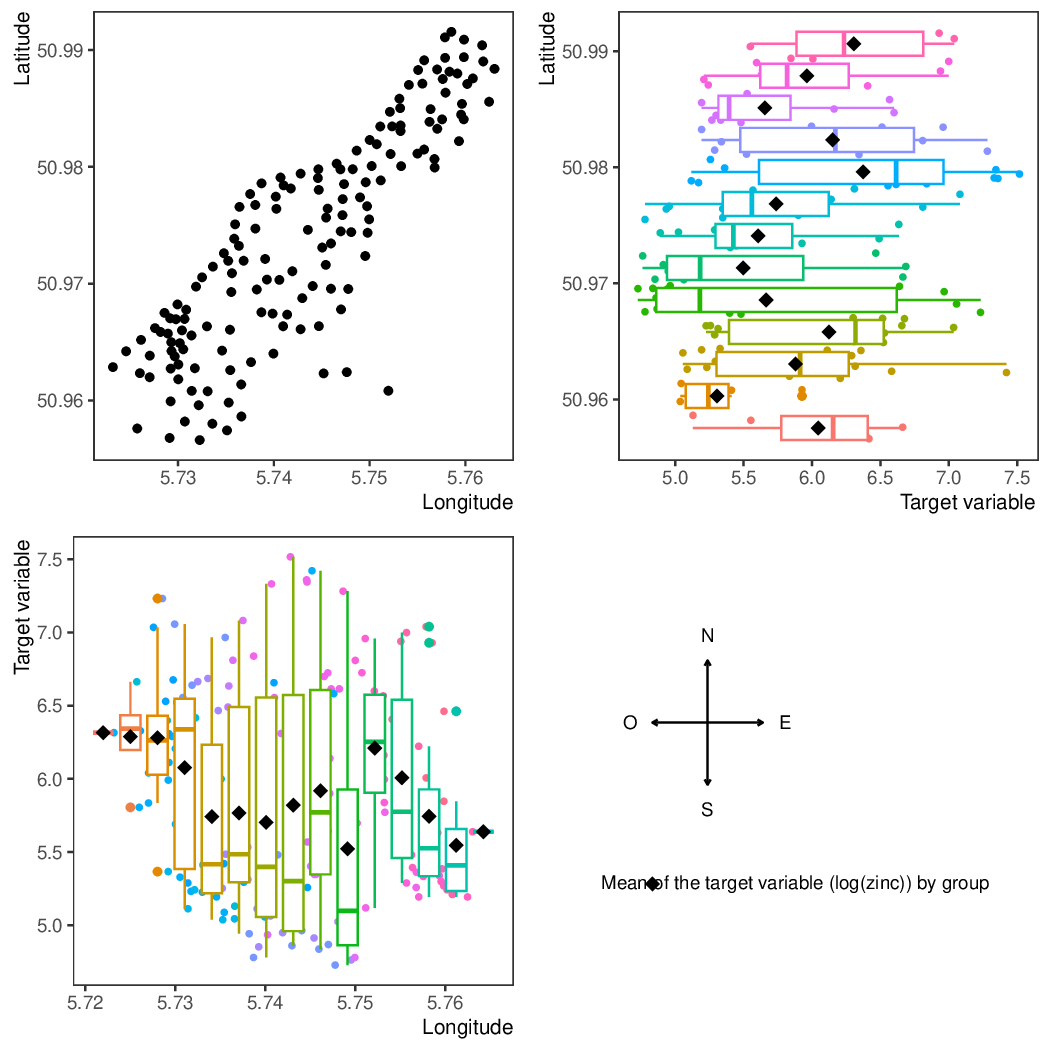}
	\caption{Spatial Distribution of the Logarithm of Zinc Concentration in the Meuse River Floodplain.}
	\label{meuse1}
\end{figure}
Figure \ref{meuse1} shows the distribution of the target variable by coordinates. In the upper left corner, the spatial distribution of the sample points is presented. The remaining two graphs illustrate the distribution of the variable's values with respect to latitude and longitude, respectively. From these plots, it can be seen that the central measures as well as their variability depend on the location of the spatial points. It is observed that the average concentration values are higher in the west compared to the east and with less dispersion. Additionally, the values tend to be higher at the north and south extremes compared to the center. As a conclusion, it seems appropriate to consider a spatial model for predicting zinc concentration at new locations.

\begin{figure}[]
	\centering
	\begin{subfigure}{0.8\textwidth}
		\includegraphics[width=\textwidth]{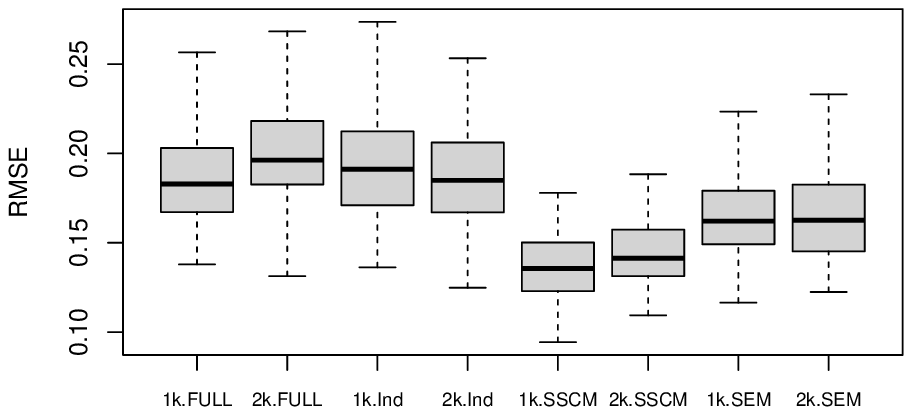}
		\caption{Setting $d=1$}
		\label{fig:Boxfirst}
	\end{subfigure}
	\hfill
	\begin{subfigure}{0.47\textwidth}
		\includegraphics[width=\textwidth]{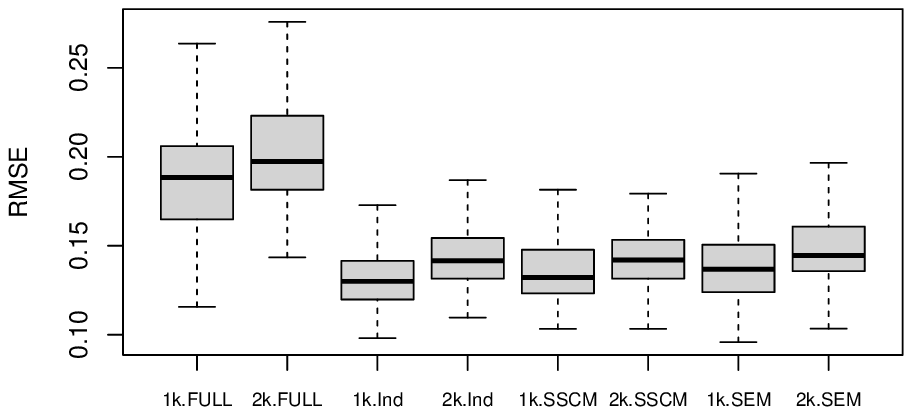}
		\caption{With $d$ optimal using Likelihood criteria}
		\label{fig:Boxsecond}
	\end{subfigure}
	\hfill
	\begin{subfigure}{0.47\textwidth}
		\includegraphics[width=\textwidth]{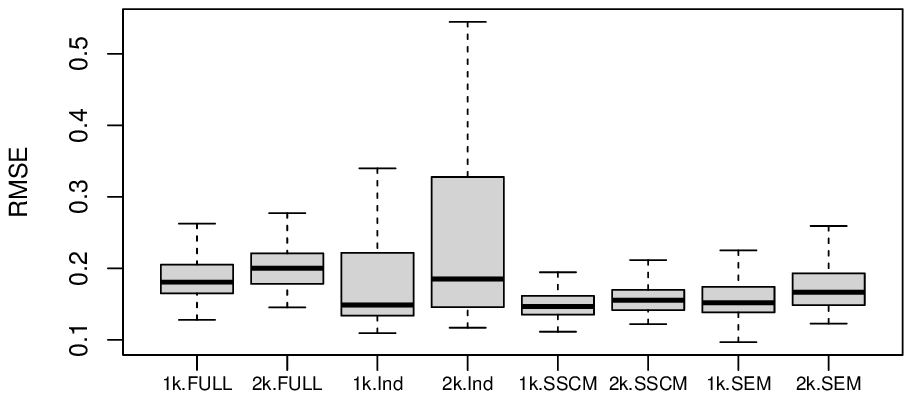}
		\caption{With $d$ optimal using CV-MPE criteria}
		\label{fig:Boxthird}
	\end{subfigure}

	\caption{Cross-validated RMSE (over $100$ replications)  for the Zinc Concentration Prediction in Meuse Data Set using SDR methods.}
	\label{meuse2}
\end{figure}

In this example, since $p$ is  relatively small, we can initially focus on the predictive performance of dimension reduction methodologies using a sing  direction ($d=1$). 

In Figure \ref{fig:Boxfirst}, we present the results of the Cross-validated Root Mean square Errors (RMSE) computed over $100$ replications using $d=1$. If we compare the nonparametric predictions using all variables (\texttt{FULL}) and the dimension reduction under independence assumption  (\texttt{Ind}), similar results are revealed, with lower average errors under \texttt{1k.FULL} and \texttt{2k.Ind}. When considering spatial dependence in the inverse models for dimension reduction, we obtain clear gains in terms of lower errors and more precise predictions. Under the SSCM, we obtain the best predictive results using the 1K predictor. In general, we do not find relevant gains from using two kernels.

{{When using likelihood-based criteria to select the dimension, Figure \ref{fig:Boxfirst} shows a significant improvement in the independent SDR over the spatial SDR methods. In this example, the LR, AIC, and BIC criteria coincide, selecting the same dimension in each iteration, resulting in an optimal $d^* = 2$ for the independent model, $d^* = 1$ for the SSCM model, and $d^* = 2$ for the SEM model. Therefore, using two directions for the independent and SEM models yields prediction errors comparable to those of the SSCM model with only one (optimal) direction.

On the other hand, under the CV-MPE criterion, the selection of $d$ varies across iterations, leading to greater variability in prediction errors, as seen in Figure \ref{fig:Boxthird}. Unlike Figure \ref{fig:Boxsecond}, this case highlights the higher variability in the RMSE of the independent model, where dimension selection is driven by prediction error rather than the likelihood function. Similar to the results when setting $d = 1$ (Figure \ref{fig:Boxfirst}), the \texttt{1k.SSCM} model demonstrates the best predictive performance.}}

\subsection{Composite Indices with Predictive Power: Proficiency Scores in Ohio Elementary Schools}
In several empirical applications of the social sciences, it is common to use composite indices that synthesize a set of characteristics to explain or predict a phenomenon of interest \citep{santeramo2015composite,yoon2018application,tomaselli2021building, khodayari2022developing, diorio2023}. The usual statistical tool for this purpose is dimension reduction methods, whether supervised (such as PFC) or unsupervised (such as PCA), choosing the projection of the first component ($d=1$) as the sought composite index; that is $CI=\sum_{i=1}^n a_i X_i= \mathbf{A}^T\mathbf{X}$, where $\mathbf{X}=(X_1,\ldots,X_p)$ are the  predictor variables (characteristics) used to construct the index and $\mathbf{A}= (a_1, \ldots,a_p)$ are their respective weights \citep{duarte2023}. In the present application, using a data set of 1965 Ohio Elementary School buildings for 2001-02 year \citep{lesage09}, the objective is to build a composite index of the characteristics of the schools of different districts in Ohio taking 25 continuous variables about building and infrastructure characteristics of the schools, teacher characteristics, spending allocation per pupil, census information of the districts and educational attainments of past students, among others. This {\it school characteristics} index is used to predict the \textit{average pupil proficiency score} ($Y$) in a district $\s_i$. Variables used to built the predictive index are:   {\it  building enrolment } ($X_1$), {\it  number of teachers} ($X_2$), {\it average years of teaching experience} ($X_3$), {\it  average teacher salary} ($X_4$), {\it pupils per teacher ratio} ($X_5$), {\it per pupil spending (PPS) on instruction} ($X_6$), {\it PPS on building operations} ($X_7$), {\it PPS on administration} ($X_8$), {\it PPS on pupil support} ($X_9$), {\it PPS on staff support} ($X_{10}$), {\it share of PPS of on instruction of total spending} ($X_{11}$), {\it share of PPS on building} ($X_{12}$), {\it share of PPS on administration} ($X_{13}$), {\it share of PPS on pupil support} ($X_{14}$), {\it share of PPS on staff support} ($X_{15}$), {\it per capita income in the area} ($X_{16}$), {\it poverty rate in the area} ($X_{17}$), {\it percent of population that is non-white} ($X_{18}$), {\it  percent of population living in same house 5 years ago} ($X_{19}$), {\it percent of population attending public schools} ($X_{20}$); educational attainment for persons 25 years and over measured by percent of {\it high-school graduates} ($X_{21}$), {\it associate degrees} ($X_{22}$), {\it college} ($X_{23}$), {\it graduate} ($X_{24}$), and {\it professional} ($X_{25}$).

In this data set we have as spatial coordinates the zip centroids of latitude and longitude having in some cases more than one school in a certain spatial point. In such cases we proceed to average the variables to have one observation per district zip code. In this way we obtain a sample of 799 spatial observations.    
\begin{figure}[]
	\centering
	\includegraphics[scale=1]{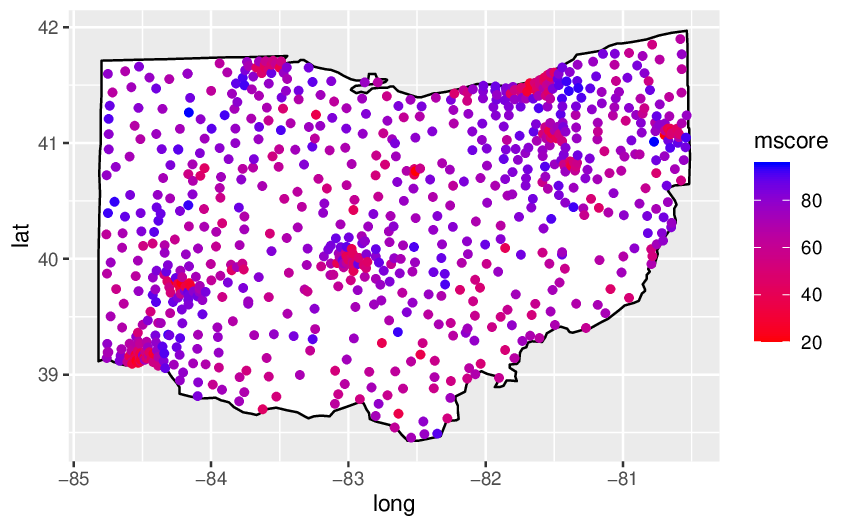}
	\caption{Mean Fourth Grade Proficiency Scores in School Districs in Ohio.}
	\label{mapohio}
\end{figure}
In Figure \ref{mapohio} we present the spatial distribution of $Y$ coloured by their values. From this can be detected some spatial clusters. In particular, we can see that some schools with the lowest mean scores are spatially concentrated in some regions such as in the south (in particular, in the south-west), north-east and in the center. In addition, near these spatial clusters, we observe abrupt changes from the lowest to the highest values in mean score, and so, clusters with extreme mean score values appears to be  neighbours. This fact could complicate the prediction based only on neighbouring values (distances), so the use of additional information is necessary to explain such close jumps in the values of the variable of interest. This additional information is taking into account with the composite index of school characteristics build it with reduction dimension techniques on the 25 predictor variables.

\begin{figure}[]
	\centering
	\includegraphics[scale=0.8]{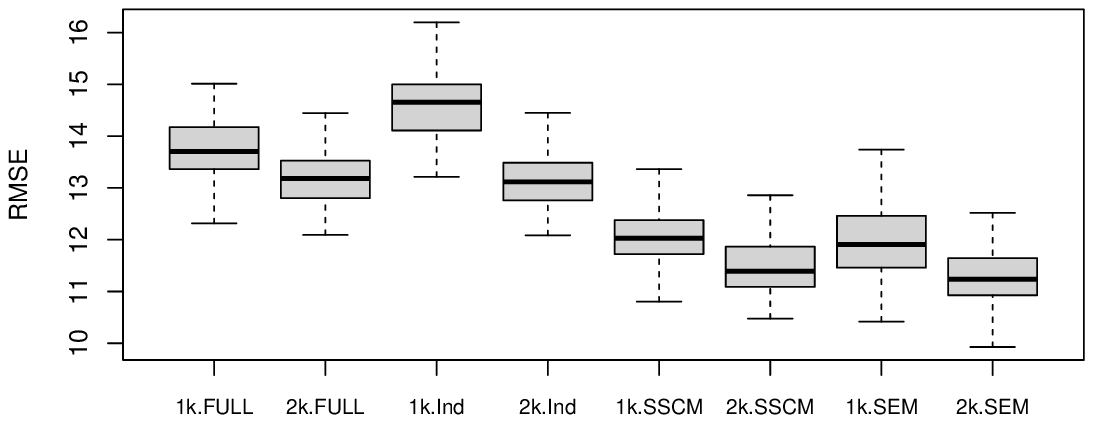}
	\caption{Cross-validated RMSE (over $100$ replications)  for the predicted Pupil Proficiency Scores in Ohio Districts Using Different Composite Indices of School Characteristics.}
	\label{bpohio}
\end{figure}

\begin{figure}[]
	\centering
	\begin{subfigure}{0.45\textwidth}
		\includegraphics[width=\textwidth]{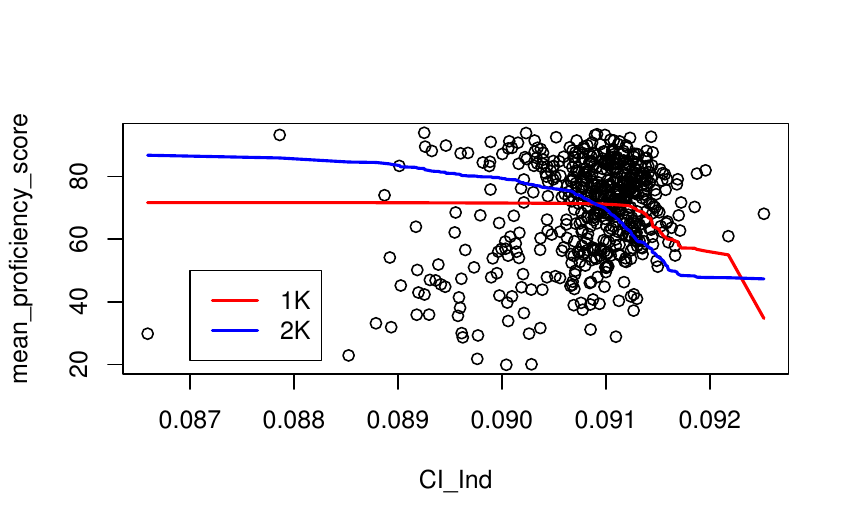}
		\caption{Under Independence}
		\label{fig:first}
	\end{subfigure}
	\hfill
	\begin{subfigure}{0.45\textwidth}
		\includegraphics[width=\textwidth]{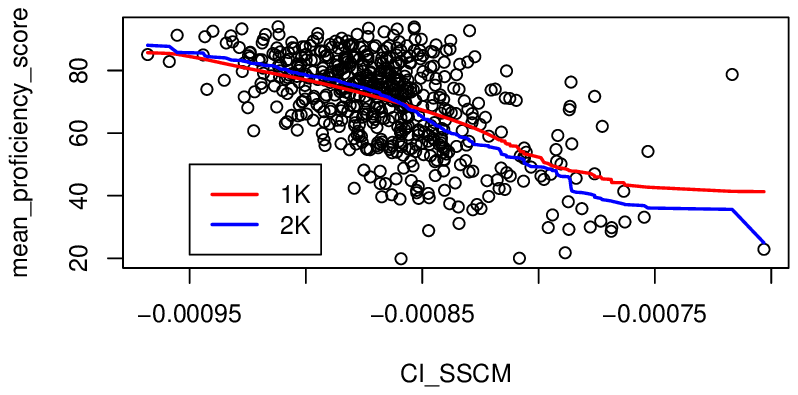}
		\caption{Under SSCM}
		\label{fig:second}
	\end{subfigure}
	\hfill
	\begin{subfigure}{0.45\textwidth}
		\includegraphics[width=\textwidth]{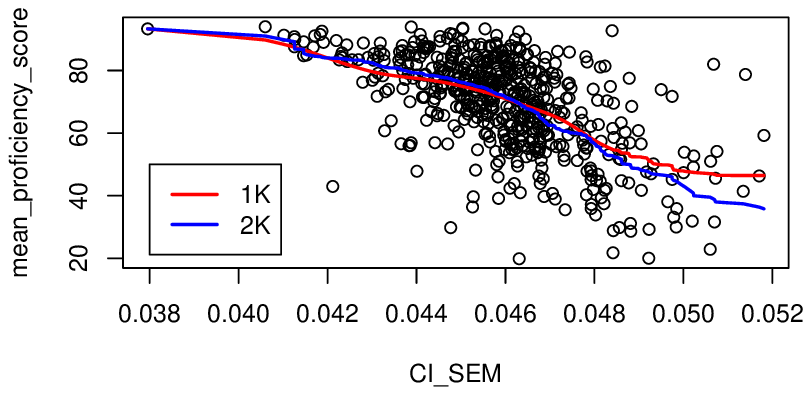}
		\caption{Under SEM}
		\label{fig:third}
	\end{subfigure}

	\caption{Nonparametric fitting in training sample using SDR under independence and  with spatial dependence.}
	\label{fig:figuresOHIO}
\end{figure}

The distribution of the Cross-validated RMSE computed over $100$ replications are visualized in Figure \ref{bpohio}. It can be seen how the indices built with reduction methods for spatial data outperform predictions obtained with full set of predictors or using dimension reduction for independent data.  In addition, and unlike the previous example, in this application can be appreciated how predictions improve when we consider the predictor with two kernels.
Figures \ref{fig:figuresOHIO} allow us to see how the inclusion of the spatial component in the reduction gives us composite indices whose relationship with the response of interest is easier to model in the training sample, and therefore we can obtain better predictors to fit the data. 
Between the index built with SDR assuming independence (Figu\-re \ref{fig:figuresOHIO}~(a)) and the one built under the SSCM model  (Figure \ref{fig:figuresOHIO}~(b)), a clear improvement is observed in the relationship to be modeled between the said index and the score. Under the SEM models  (Figure \ref{fig:figuresOHIO}~(c)), a further improvement is observed, and this is revealed in lower predictive errors on the test sample.

\subsection{Economic Growth Rates Prediction in the World}
In this example the aim is to predict the average Gross Domestic Product (GDP) rate of a country using several predictors about economic, social, demographic and cultural characteristics  of a country. There is a vast literature that empirically shows the existence of a significant spatial dependence on economic growth among countries \citep[e.g.][among others]{bosker2009economic, Amidi20, Chih22, mahran2023impact}. In general, these studies show that in addition to the economic, social, institutional, and geographical characteristics of a country, its own economic growth is affected by the growth rates of its neighbouring countries. Therefore, such spatial effects must be taken into account in the modelling both for the estimation of parameters (effects of covariates) of interest and for the task of predicting the average GDP growth (that is, our response variable $Y$).

We use a data set of 72 countries that covers their growth rates over 1960-80 period taken from Fernandez, Ley and Steel (2001). For these 72 countries we have 19 continuous predictor variables:  {\it GDP level in 1960} ($X_1$), {\it Life expectancy} ($X_2$), {\it Primary school enrollment in 1960} ($X_3$), {\it Fraction GDP in mining} ($X_4$), {\it Degree of capitalism} ($X_5$), {\it Number of years open economy} ($X_6$), {\it Fraction speaking a foreign language} ($X_7$), {\it Exchange rate distortions} ($X_8$), {\it Equipment investment} ($X_9$), {\it non-equipment investment} ($X_{10}$), {\it Public education share} ($X_{11}$), {\it Civil liberties} ($X_{12}$), {\it Absolute latitude} ($X_{13}$), {\it Ethnolinguistic fractionalization} ($X_{14}$), {\it Fraction Muslim} ($X_{15}$), {\it Fraction Protestant} ($X_{16}$), {\it Fraction Catholic} ($X_{17}$), {\it Primary exports} ($X_{18}$)and the {\it Ratio of workers to the population} ($X_{19}$).
\begin{figure}[!ht]
	\centering
	\includegraphics[scale=0.5]{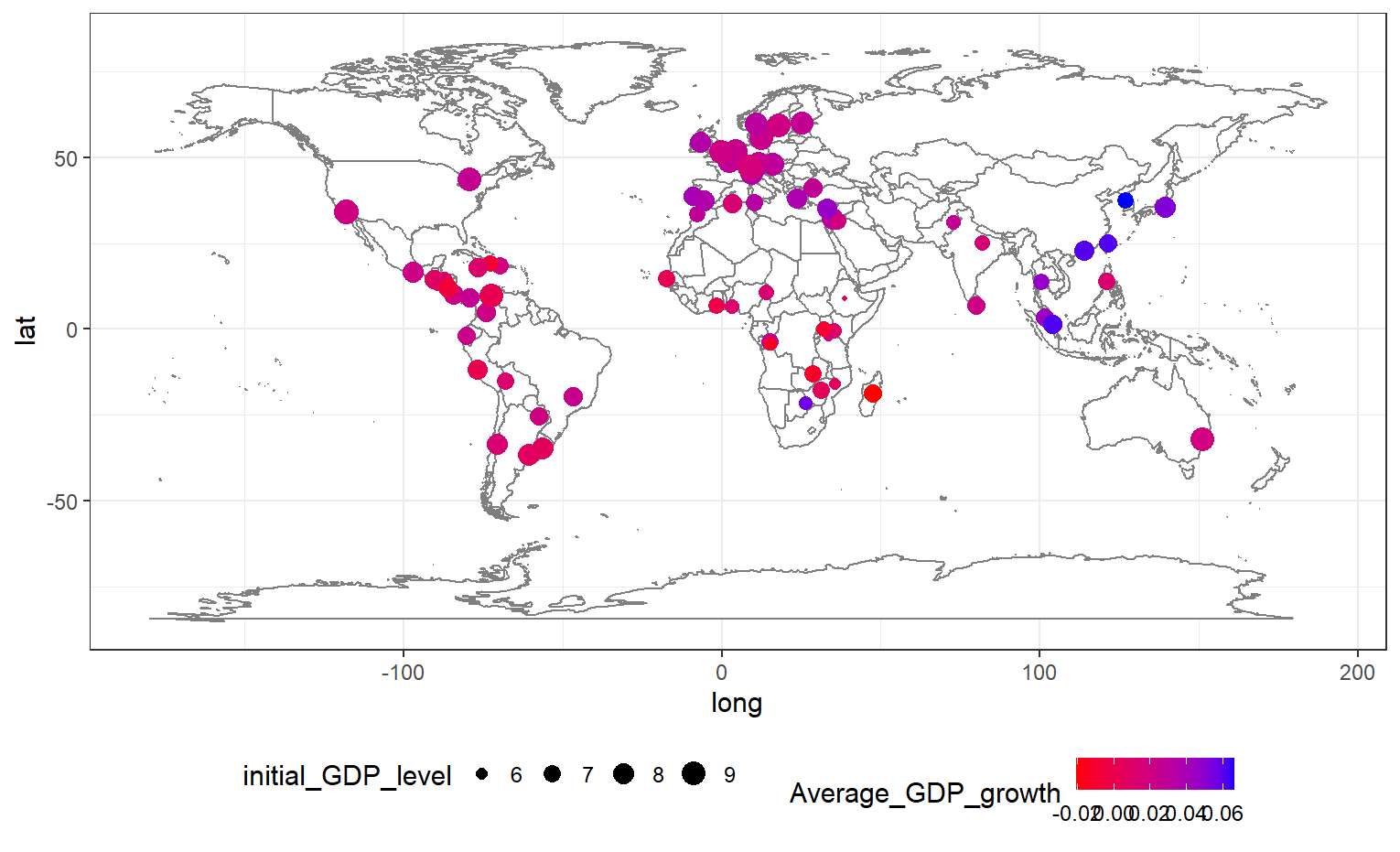}
	\caption{Average Gross Domestic Product (GDP) 1960-80 for 72 countries and their initial GPD level.}
	\label{WorldGDP}
\end{figure}

Figure \ref{WorldGDP} illustrates the values of the response variable $Y$  (the average GDP growth), for each of the 72 selected countries, differentiated by color. Additionally, the size of the points indicates the initial GDP level in 1960, which is included in the covariate vector. This figure highlights various spatial patterns in countries
economic growth. For example, for Southeast Asia countries it is observed high growth starting from a low initial GDP, whereas in developed economies of Europe, North America and Oceania a medium-high GDP growth is observed in the period but with lower rates than the first ones. This is in line with the so-called {\it Beta-convergence} \citep{barro1992convergence} in the economic development literature, where economies with low per capita GDP tends to grow faster than high-income countries. However, for most African countries it is observed low initial GDP and growth, whereas South America is characterized by low growth with medium-high initial GDP. Therefore, some clear spatial clusters in growth patterns can be observed, which motivates the inclusion of spatial correlations in modeling GDP growth of countries at a global level to improve prediction.

\begin{table}[!ht]
\centering

\begin{tabular}{rcccc}
\hline
           &         \multicolumn{ 4}{c}{$d$ Selection Criteria} \\
\hline
    Method &         LR &        AIC &        BIC &         CV \\
\hline
  \texttt{ 1k.FULL} &     1.6437 &     1.6762 &     1.7025 &     1.6935 \\

           &     (0.2384) &     (0.2651) &     (0.2419) &     (0.2731) \\

           &      ---      &     ---       &     ---       &    ---        \\
\hline
     \texttt{ 2k.FULL} &     1.5662 &     1.5785 &     1.6301 &     1.5832 \\

           &     (0.3321) &      (0.3550) &     (0.3038) &     (0.2924) \\

           &       ---     &     ---       &     ---       &   ---         \\
\hline
      \texttt{ 1k.Ind} &     1.3650 &     1.3624 &       1.3900 &     1.5317 \\

           &     (0.3137) &     (0.3429) &     (0.3677) &     (0.3368) \\

           &         [ 2] &          [ 2] &         [ 2] &          [ 1] \\
\hline
          \texttt{ 2k.Ind} &     1.3755 &     1.3343 &     1.3881 &     1.3211 \\

           &     (0.3385) &     (0.3057) &      (0.3140) &     (0.3001) \\

           &          [ 2] &          [ 2] &          [ 2] &          [ 2] \\
\hline
         \texttt{ 1k.SSCM} &     1.3772 &     1.4492 &     1.4932 &     1.4593 \\

           &     (0.3113) &     (0.3084) &     (0.3343) &     (0.3252) \\

           &          [ 3] &          [ 3] &          [ 1] &          [ 2] \\
\hline
   \texttt{ 2k.SSCM} &      1.3450 &     1.3904 &     1.4044 &     1.3274 \\

           &     (0.3182) &     (0.3061) &     (0.3431) &     (0.3975) \\

           &          [ 3] &          [ 3] &          [ 1] &          [ 2] \\
\hline
    \texttt{ 1k.SEM} &     1.3002 &     1.3081 &     1.2644 &     1.4561 \\

           &     (0.2612) &     (0.2764) &     (0.2119) &     (0.2255) \\

           &          [ 3] &          [ 2] &          [ 2] &          [ 1] \\
\hline
   \texttt{ 2k.SEM}  & {\bf 1.2588} & {\bf 1.2685} & {\bf 1.2237} & {\bf 1.2665} \\

           &     (0.2489) &     (0.2919) &     (0.2469) &     (0.2557) \\

           &      [ 3]     &          [ 2] &          [ 2] &          [ 2] \\
\hline
\end{tabular}  
	\caption{Average cross-validated RMSE (computed over 100 replications) for the spatial prediction of average GDP growth across countries. Standard deviations are reported in parentheses, and the median of optimal $d^*$ is shown in brackets.}
 \label{tabgdperror}
\end{table}
{{In Table \ref{tabgdperror}, we present the average cross-validated RMSE and  its standard error, computed over 100 replications. The table also includes the median of the optimal dimension, $d^*$, selected according to different criteria. All the analysis were conducted for both non-parametric predictors. The table demonstrates that all dimension reduction methodologies yield better predictive performance compared to models utilizing the full set of covariates without any reduction. Although the SSCM model outperforms the independence model in only two cases, the SEM model consistently delivers the best predictive results across all values of $d$ and predictive rules, as emphasized in bold in the table. In general, all methods yield better predictions when using the two-kernel predictor. 

Regarding the selection of $d$, we observe that for the independent case, all parametric criteria consistently select $d^*=2$. This may be due to the fact that the AIC, BIC, and LR criteria do not depend on the predictive rule. The CV-MPE criterion, however, selects a lower dimension when using just one kernel in the predictor. Under the SSCM model, both AIC and LR select the same dimension ($d^*=3$), while for the SEM model, AIC and BIC selected the same dimension, which is $d^*=2$. For these two spatial models, the CV-MPE criterion identifies different dimensions only in the case of the SEM model. In conclusion,  the CV-MPE criterion tends to select e lower dimensions than the parametric criteria and also chooses different dimensions depending on the prediction method used.}}

\section{Discussion}\label{discussion}
\
In this section, we discuss the application of our proposed sufficient dimension reduction methodologies for spatial prediction, highlighting the results obtained from experiments conducted with real data. As emphasized, the predictive gains from dimensionality reduction compared to using all variables without reduction are substantial. However, a more detailed comparison of the dimension reduction methodologies is necessary, particularly between the SSCM and SEM models proposed for spatial data.

The primary difference between  SSCM and SEM  lies in the specification of the spatial association matrix $\Sb$. The specifications of SSCM and SEM through $\Hb$ and $\Wbf_{\theta}$ correspond to two alternative perspectives, namely, the geostatistic and lattice approaches, respectively \citep{cressie2015statistics}. In the geostatistical approach,  spatial variation  is considered a continuous process, meaning data is indexed continuously within the domain $D$  and the spatial covariance (and/or variogram) is modeled as a smooth function (e.g., exponential or spherical) of the distance between any two locations. In contrast, the lattice approach considers spatial covariance as arising from interactions among discrete spatial objects, defining a neighborhood structure using a spatial weights matrix \citep{anselin2001spatial, zhang2014scale}.\\
In this way, with  both methodologies we seek to make SDR extensive for various applications that involve the spatial dependency, considering  that  geostatistical data are more frequent in environmental and geological studies, whereas the lattice data  are more common in economics, geography and regional sciences. By applying both spatial SDR methods to both types of data, it is important to consider the feasibility of doing this and how the results may be affected when the method does not align with the data type. As suggest by \cite{cressie2015statistics}, methods from one class of problems can be adapted for use in another class. Geostatistical models can be adjusted to properly handle lattice data by assuming that the values of the variable of interest on a lattice occur at the centers of the lattice units (centroids), thereby treating $D$ as continuous to permit spatial prediction at intermediate locations \citep{zhang2014scale}. However,  it must be considered that using the centers or centroids of the regions could influence the distance measurement and potentially modify the existing spatial dependence structure in the data \citep{zhao2004investigating}. Lattice models can be used for geostatistical data by aggregating the latter into areas to generate lattice data \citep{zhang2014scale}. Considering we have neighborhood matrices for lattice data and distance matrices for geostatistical data, it is possible to convert from one to the other by setting, for example, a minimum distance beyond which observations are no longer considered neighbors \citep{de2018codifying}. In fact, there are several approaches to construct the spatial weights matrix. By adopting a distance-based matrix, we seek to 'unify' methodologies and be consistent with non-parametric prediction rules based on spatial distances. 

In our real data applications, we can see that for the zinc prediction in Meuse river, where the physical features of spatial units can measure in a continuous domain, the SSCM  outperform the SEM method. Therefore, as we could have hypothesized, for data of a geostatistical nature the SSCM gives better results.\\
The other two application involve lattice data but exhibit very different spatial characteristics. For the case of proficiency scores prediction in Ohio elementary school, we encountered an increased density of spatially clustered points in certain areas, making the geostatistical distance approach more suitable for treating this lattice data with ZIP code centroids. This could explain why the results with SSCM method are comparable with those obtained with the SEM method, although, as we demonstrated, reduction with the SEM is the best alternative.\\
In contrast, for the third application, the spatial coordinates are the capital cities of countries around the world. Although the distance between capital cities  account for the geographic distribution of the population of each nation \citep{Amidi20}, for large states and irregular territories, the geometric distance between these spatial coordinates may overstate the actual distances across state boundaries \cite{gleditsch2001measuring}. This may explain the poor performance of SSCM, even when compared to SDR under independence assumption. In fact, as the SDR estimates under SEM  depend on the choice of the weights matrix, a different criterion (e.g., based on contiguity or $k$-nearest neighbors) could be a more suitable option for this case. However, even using the distance-based matrix, we observe the superiority of SEM for these lattice data. Nevertheless, studying the effect of different approaches for weight matrix construction could be an interesting empirical experiment for future research and a possible extension for related applications.

\section{Conclusions}\label{conclus}

We presented a sufficient dimension reduction (SDR) methodology for spatial data, evaluating their predictive performance using two types of spatial non-parametric regression models. The  SDR  for the regression of  $Y_{\s}$ on $\X_{s}$ emerged from an inverse regression model where  the errors follow a multivariate normal distribution and are spatially associated through a positive definite matrix $\Sb$.
The elements of $\Sb$ represent the spatial association between any pair of errors measured at different locations. Depending on the specification of $\Sb$, we obtain either a Spatial Separable Covariance Model (SSCM) or a Spatial autoregressive Error Model (SEM), deriving their respective maximum likelihood estimates for the SDR.

Leveraging the advantages of dimensionality reduction, we propose two non-parametric regression functions as prediction rules: the classic Nadaraya-Watson estimator with a kernel evaluated on the distance of the reduced predictor, and another using two kernels that include the distance between locations.

From simulations and real data examples, we demonstrate that our spatial SDR methods outperform the non-reduction strategy and surpass the analogous reduction method that assumes independence; that is, the Principal Fitted Components (PFC). Furthermore, incorporating a kernel that accounts for spatial distance into the prediction rule results in substantial predictive improvements.

\section*{Acknowledgements}

This paper have been supported by the MATH-AmSud international grant \textit{22MATH-07} and by the national grants \textit{CAID 503-20190-100022LI}, \textit{CAID UNL 506-20190-100083LI},\textit{PICT ANCyT 2018-03005} and \textit{PIP CONICET 11220200101595CO}. We appreciate the support and funding provided by this initiative, which has been crucial in the completion of this work.

\bibliographystyle{elsarticle-harv}

\begin{thebibliography}{}

\bibitem[Adragni and Cook, 2009]{AC09}
Adragni, K. and Cook, R. (2009).
\newblock Sufficient dimension reduction and prediction in regression.
\newblock {\em Philosophical Transactions of the Royal Society of London A: Mathematical, Physical and Engineering Sciences}, 367(1906):4385--4405.

\bibitem[Amidi and Majidi, 2020]{Amidi20}
Amidi, S. and Majidi, A.~F. (2020).
\newblock Geographic proximity, trade and economic growth: a spatial econometrics approach.
\newblock {\em Annals of GIS}, 26(1):49--63.

\bibitem[Anselin, 1988]{Anselin88}
Anselin, L. (1988).
\newblock {\em Spatial Econometrics: Methods and Models}.
\newblock Springer.

\bibitem[Anselin, 2001]{anselin2001spatial}
Anselin, L. (2001).
\newblock Spatial effects in econometric practice in environmental and resource economics.
\newblock {\em American Journal of Agricultural Economics}, 83(3):705--710.

\bibitem[Barbarino and Bura, 2015]{Barbarino2015}
Barbarino, A. and Bura, E. (2015).
\newblock {Forecasting with Sufficient Dimension Reductions}.
\newblock Finance and Economics Discussion Series 2015-74, Board of Governors of the Federal Reserve System (U.S.).

\bibitem[Barbarino and Bura, 2024]{BARBARINO20241}
Barbarino, A. and Bura, E. (2024).
\newblock Forecasting near-equivalence of linear dimension reduction methods in large panels of macro-variables.
\newblock {\em Econometrics and Statistics}, 31:1--18.

\bibitem[Barro and Sala-i Martin, 1992]{barro1992convergence}
Barro, R.~J. and Sala-i Martin, X. (1992).
\newblock Convergence.
\newblock {\em Journal of political Economy}, 100(2):223--251.

\bibitem[Biau and Cadre, 2004]{Biau2004}
Biau, G. and Cadre, B. (2004).
\newblock Nonparametric spatial prediction.
\newblock {\em Statistical Inference for Stochastic Processes}, 7(3):327--349.

\bibitem[Bosker and Garretsen, 2009]{bosker2009economic}
Bosker, M. and Garretsen, H. (2009).
\newblock Economic development and the geography of institutions.
\newblock {\em Journal of economic geography}, 9(3):295--328.

\bibitem[Bura et~al., 2016]{BDF16}
Bura, E., Duarte, S., and Forzani, L. (2016).
\newblock Sufficient reductions in regressions with exponential family inverse predictors.
\newblock {\em Journal of the American Statistical Association}, 111(515):1313--1329.

\bibitem[Bura et~al., 2022]{bura22}
Bura, E., Forzani, L., Garc\'ia-Arancibia, R., Llop, P., and Tomassi, D. (2022).
\newblock Suffcient reductions in regression with mixed predictors.
\newblock {\em Journal of Machine Learning Research}, 23:1--47.

\bibitem[Casella and Berger, 1990]{casellaberger90}
Casella, G. and Berger, R. (1990).
\newblock {\em Statistical Inference}.
\newblock The Wadsworth \& Brooks/Cole Statistics/Probability Series.

\bibitem[Chih et~al., 2022]{Chih22}
Chih, Y.-Y., Kishan, R.~P., and Ojede, A. (2022).
\newblock Be good to thy neighbours: A spatial analysis of foreign direct investment and economic growth in sub-saharan africa.
\newblock {\em The World Economy}, 45(3):657--701.

\bibitem[Cook, 1998]{cook98}
Cook, R. (1998).
\newblock {\em Regression Graphics}.
\newblock Wiley, New York.

\bibitem[Cook and Weisberg, 1991]{CookWeisberg1991}
Cook, R. and Weisberg, S. (1991).
\newblock Discussion of sliced inverse regression for dimension reduction.
\newblock {\em Journal of the American Statistical Association}, 86:328--332.

\bibitem[Cook, 2007]{Cook2007}
Cook, R.~D. (2007).
\newblock Fisher lecture: {D}imension reduction in regression.
\newblock {\em Statist. Sci.}, 22:1--26.

\bibitem[Cook and Forzani, 2008]{CookForzani2008}
Cook, R.~D. and Forzani, L. (2008).
\newblock Principal fitted components for dimension reduction in regression.
\newblock {\em Statist. Sci.}, 23:485--501.

\bibitem[Cressie, 2015]{cressie2015statistics}
Cressie, N. (2015).
\newblock {\em Statistics for spatial data}.
\newblock John Wiley \& Sons.

\bibitem[Dabo-Niang et~al., 2016]{DN16}
Dabo-Niang, S., Ternynck, C., and Yao, A. (2016).
\newblock Nonparametric prediction of spatial multivariate data.
\newblock {\em Journal of Nonparametric Statistics}, 28(2):428--458.

\bibitem[Dabo-Niang and Yao, 2007]{DaboNiang2007}
Dabo-Niang, S. and Yao, A.~F. (2007).
\newblock Kernel regression estimation for continuous spatial processes.
\newblock {\em Math. Methods Statist.}, 16:298--317.

\bibitem[Datta and Loh, 2022]{datta2022}
Datta, S. and Loh, J. (2022).
\newblock Sufficient dimension reduction for spatial point processes using weighted principal support vector machines.
\newblock {\em Statistics and Its Interface}, 15(4):415--431.

\bibitem[De~Bellefon et~al., 2018]{de2018codifying}
De~Bellefon, M., Loonis, V., and Le~Gleut, R. (2018).
\newblock Codifying the neighbourhood structure.
\newblock {\em Handbook of Spatial Analysis-Theory and practical application with R. INSEE Eurostat}, pages 31--47.

\bibitem[Duarte et~al., 2023]{duarte2023}
Duarte, S., Forzani, L., Garc{\'\i}a~Arancibia, R., Llop, P., and Tomassi, D. (2023).
\newblock Socioeconomic index for income and poverty prediction: A sufficient dimension reduction approach.
\newblock {\em Review of Income and Wealth}, 69:318--346.

\bibitem[D’Iorio et~al., 2023]{diorio2023}
D’Iorio, S., Forzani, L., Garc{\'\i}a~Arancibia, R., and Girela, I. (2023).
\newblock Predictive power of composite socioeconomic indices in regression and classification: Principal components and partial least squares.
\newblock Technical report, Working Paper 246, Red Nacional de Investigadores en Economía (RedNIE).

\bibitem[Forzani et~al., 2018]{forzani18}
Forzani, L., Garc\'ia-Arancibia, R., Llop, P., and Tomassi, D. (2018).
\newblock Supervised dimension reduction for ordinal predictors.
\newblock {\em Computational Statistics and Data Analysis}, 125:136--155.

\bibitem[Gleditsch and Ward, 2001]{gleditsch2001measuring}
Gleditsch, K.~S. and Ward, M.~D. (2001).
\newblock Measuring space: A minimum-distance database and applications to international studies.
\newblock {\em Journal of Peace Research}, 38(6):739--758.

\bibitem[Goulard et~al., 2017]{goulard2017}
Goulard, M., Laurent, T., and Thomas-Agnan, C. (2017).
\newblock About predictions in spatial autoregressive models: Optimal and almost optimal strategies.
\newblock {\em Spatial Economic Analysis}, 12(2-3):304--325.

\bibitem[Guan, 2008]{guan08}
Guan, Y. (2008).
\newblock On consistent nonparametric intensity estimation for inhomogeneous spatial point processes.
\newblock {\em Journal of the American Statistical Association}, 103(483):1238--124.

\bibitem[Guan and Wang, 2010]{guan10}
Guan, Y. and Wang, H. (2010).
\newblock Sufficient dimension reduction for spatial point processes directed by gaussian random fields.
\newblock {\em Journal Of The Royal Statistical Society, Series B}, 72(3):367--387.

\bibitem[Heaton et~al., 2019]{heaton2019case}
Heaton, M.~J., Datta, A., Finley, A.~O., Furrer, R., Guinness, J., Guhaniyogi, R., Gerber, F., Gramacy, R.~B., Hammerling, D., Katzfuss, M., et~al. (2019).
\newblock A case study competition among methods for analyzing large spatial data.
\newblock {\em Journal of Agricultural, Biological and Environmental Statistics}, 24:398--425.

\bibitem[Hu et~al., 2009]{hu2009spatial}
Hu, J., Kaparias, I., and Bell, M.~G. (2009).
\newblock Spatial econometrics models for congestion prediction with in-vehicle route guidance.
\newblock {\em IET Intelligent Transport Systems}, 3(2):159--167.

\bibitem[Junttila and Laine, 2017]{junttila2017bayesian}
Junttila, V. and Laine, M. (2017).
\newblock Bayesian principal component regression model with spatial effects for forest inventory variables under small field sample size.
\newblock {\em Remote Sensing of Environment}, 192:45--57.

\bibitem[Khodayari~Moez et~al., 2022]{khodayari2022developing}
Khodayari~Moez, E., Maximova, K., Sim, S., Senthilselvan, A., and Pabayo, R. (2022).
\newblock Developing a socioeconomic status index for chronic disease prevention research in canada.
\newblock {\em International Journal of Environmental Research and Public Health}, 19(13):7800.

\bibitem[Kopczewska, 2023]{Kop23}
Kopczewska, K. (2023).
\newblock Spatial bootstrapped microeconometrics: Forecasting for out-of-sample geo-locations in big data.
\newblock {\em Scandinavian Journal of Statistics}, 50(3):1391--1419.

\bibitem[LeSage and Pace, 2009]{lesage09}
LeSage, J. and Pace, R. (2009).
\newblock {\em Introduction to Spatial Econometrics, 1st Edition}.
\newblock Chapman and Hall/CRC.

\bibitem[Li and Wang, 2007]{LiWang2007}
Li, B. and Wang, S. (2007).
\newblock {On directional regression for dimension reduction}.
\newblock {\em Journal of the American Statistical Association}, 102(479):997--1008.

\bibitem[Li, 1991]{Li1991}
Li, K. (1991).
\newblock Sliced inverse regression for dimension reduction (with discussion).
\newblock {\em Journal of the American Statistical Association}, 86(6):316--342.

\bibitem[Loubes and Yao, 2013]{loubes13}
Loubes, J. and Yao, A. (2013).
\newblock Kernel inverse regression for random fields.
\newblock {\em International Journal of Applied Mathematics and Statistics}, 32:1--26.

\bibitem[Mahran, 2023]{mahran2023impact}
Mahran, H.~A. (2023).
\newblock The impact of governance on economic growth: spatial econometric approach.
\newblock {\em Review of Economics and Political Science}, 8(1):37--53.

\bibitem[Matilainen et~al., 2019]{matilainen2019sliced}
Matilainen, M., Croux, C., Nordhausen, K., and Oja, H. (2019).
\newblock Sliced average variance estimation for multivariate time series.
\newblock {\em Statistics}, 53(3):630--655.

\bibitem[May and {Moradi Rekabdarkolaee}, 2024]{MAY2024100838}
May, P. and {Moradi Rekabdarkolaee}, H. (2024).
\newblock Dimension reduction for spatial regression: Spatial predictor envelope.
\newblock {\em Spatial Statistics}, 61:100838.

\bibitem[Menezes et~al., 2010]{menezes10}
Menezes, R., Garc\'ia-Soid\'an, P., and Ferreira, C. (2010).
\newblock Nonparametric spatial prediction under stochastic sampling design.
\newblock {\em J. Nonparametr. Stat.}, 22:363--377.

\bibitem[Meyer and Pebesma, 2021]{meyer2021predicting}
Meyer, H. and Pebesma, E. (2021).
\newblock Predicting into unknown space? estimating the area of applicability of spatial prediction models.
\newblock {\em Methods in Ecology and Evolution}, 12(9):1620--1633.

\bibitem[Nikparvar and Thill, 2021]{Nikparvar21}
Nikparvar, B. and Thill, J.-C. (2021).
\newblock Machine learning of spatial data.
\newblock {\em ISPRS International Journal of Geo-Information}, 10(9).

\bibitem[Reinsel and Velu, 1998]{ReinselVelu98}
Reinsel, G. and Velu, R. (1998).
\newblock {\em Multivariate Reduced-Rank Regression: Theory and Applications}.
\newblock Springer.

\bibitem[Sampson et~al., 2013]{SAMPSON2013383}
Sampson, P.~D., Richards, M., Szpiro, A.~A., Bergen, S., Sheppard, L., Larson, T.~V., and Kaufman, J.~D. (2013).
\newblock A regionalized national universal kriging model using partial least squares regression for estimating annual pm2.5 concentrations in epidemiology.
\newblock {\em Atmospheric Environment}, 75:383--392.

\bibitem[Santeramo, 2015]{santeramo2015composite}
Santeramo, F.~G. (2015).
\newblock On the composite indicators for food security: decisions matter!
\newblock {\em Food Reviews International}, 31(1):63--73.

\bibitem[Tomaselli et~al., 2021]{tomaselli2021building}
Tomaselli, V., Fordellone, M., and Vichi, M. (2021).
\newblock Building well-being composite indicator for micro-territorial areas through pls-sem and k-means approach.
\newblock {\em Social Indicators Research}, 153:407--429.

\bibitem[Yoon and Klasen, 2018]{yoon2018application}
Yoon, J. and Klasen, S. (2018).
\newblock An application of partial least squares to the construction of the social institutions and gender index (sigi) and the corruption perception index (cpi).
\newblock {\em Social Indicators Research}, 138:61--88.

\bibitem[Zhang et~al., 2014]{zhang2014scale}
Zhang, J., Atkinson, P., and Goodchild, M.~F. (2014).
\newblock {\em Scale in spatial information and analysis}.
\newblock CRC Press.

\bibitem[Zhao and Wall, 2004]{zhao2004investigating}
Zhao, Y. and Wall, M.~M. (2004).
\newblock Investigating the use of the variogram for lattice data.
\newblock {\em Journal of Computational and Graphical Statistics}, 13(3):719--738.

\end{thebibliography}


\newpage

\appendix


\section{Proofs of Section \ref{secSSCM} }\label{AppendixA}

\begin{proofprop}{SSCM-estimators}
Assuming that the correlation matrix $\Hbf$ is given, the next step is to estimate the mean $\mubf$. Taking derivative with respect to $\mubf$ in \eqref{log-likelihood-SSCM}, except for some terms  independent of $\mubf$ we have that,
\begin{align*}
\frac{\partial l(\Xbb|\Y,  \Hb;\mubf,\Ab, \Bb, \Deltabf)}{\partial \mubf} &=\\ &\hspace{-2cm}=  \frac{\partial}{\partial \mubf}\left\{-  \frac{1}{2} \tr\Big( \Hb^{-1/2} (\Xbb -  \ones_n  \mubf^T - \Fbb \Bb^T \Ab^T) \Deltabf^{-1}  (\Xbb -  \ones_n  \mubf^T - \Fbb \Bb^T \Ab^T)^T  \Hb^{-1/2}\Big) \right\} \\ &\hspace{-2cm}= \Deltabf^{-1} (\Xbb^T -  \mubf \ones_n ^T - \Ab\Bb \Fbb^T  ) \Hb^{-1} \ones_n,
\end{align*}
which is 0  if and only if 
\begin{equation}\label{mu-SSCM}
    \widetilde \mubf  = (\Xbb^T - \Ab \Bb \Fbb^T)\Hb^{-1}\ones_n (\ones_n^T\Hb^{-1}\ones_n)^{-1}.
\end{equation}
Therefore, in \eqref{log-likelihood-SSCM} we have that
\begin{align*}
	\Xbb -  \ones_n  \widehat \mubf^T - \Fbb \Bb^T \Ab^T &=\\ &\hspace{-1cm}= \Xbb -  (\ones_n^T\Hb^{-1}\ones_n)^{-1} \ones_n  \ones_n^T \Hb^{-1}  \Xbb - (\ones_n^T\Hb^{-1}\ones_n)^{-1} \ones_n  \ones_n^T \Hb^{-1} \Fbb \Bb^T \Ab^T - \Fbb \Bb^T \Ab^T \\ &\hspace{-1cm}= (\ones_n - (\ones_n^T\Hb^{-1}\ones_n)^{-1} \ones_n  \ones_n^T \Hb^{-1})\Xbb - (\ones_n - (\ones_n^T\Hb^{-1}\ones_n)^{-1} \ones_n  \ones_n^T \Hb^{-1}) \Fbb \Bb^T \Ab^T \\ &\hspace{-1cm} \doteq   \Hb^c\Xbb -  \Hb^c\Fbb \Bb^T \Ab^T,
\end{align*}
where $\Hb^c\doteq \ones_n - (\ones_n^T\Hb^{-1}\ones_n)^{-1} \ones_n  \ones_n^T \Hb^{-1}$, so that
\[ 
\Hb^{-1/2}  (\Xbb -  \ones_n  \widehat \mubf^T - \Fbb \Bb^T \Ab^T) =  \Hb^{-1/2} \Hb^c\Xbb - \Hb^{-1/2} \Hb^c\Fbb \Bb^T \Ab^T.
\]
Defining $\Xbbb$ and $\Fbbb$ as  
\[
\Xbbb =\Hb^{-1/2} \Hb^c\Xbb \qquad \text{ and } \qquad \Fbbb = \Hb^{-1/2} \Hb^c\Fbb,
\]
we have that, apart from irrelevant constants, the log-likelihood \eqref{log-likelihood-SSCM} becomes
\begin{align}\label{log-likelihood-SSCM-1}
	l(\Xbb|\Y, \Hb;\Ab, \Bb, \Deltabf)  &= - \frac{p}{2} \log | \Hb|  - \frac{n}{2} \log |\Deltabf| -  \frac{1}{2} \tr\Big( (\Xbbb - \Fbbb \Bb^T \Ab^T) \Deltabf^{-1} (\Xbbb- \Fbbb \Bb^T \Ab^T )^T \Big).
\end{align}
Now, we take derivative with respect to $\Deltabf$ to get
\begin{align*}
	\frac{\partial l(\Xbb|\Y, \Hb;\mubf,\Ab, \Bb, \Deltabf)}{\partial \Deltabf^{-1}}&=  \frac{\partial}{\partial \Deltabf^{-1}} \left\{\frac{n}{2} \log |\Deltabf^{-1}| -  \frac{1}{2} \tr\Big((\Xbbb - \Fbbb \Bb^T \Ab^T ) \Deltabf^{-1} (\Xbbb- \Fbbb \Bb^T \Ab^T )^T \Big)\right\} \\ &= \frac{n}{2}\Deltabf -  \frac{1}{2} (\Xbbb - \Fbbb \Bb^T \Ab^T )^T (\Xbbb - \Fbbb \Bb^T \Ab^T),
\end{align*}
which is 0  if and only if 
\begin{equation}\label{Delta-SSCM}
	\widetilde \Deltabf =  \frac{1}{n}(\Xbbb - \Fbbb \Bb^T \Ab^T )^T (\Xbbb - \Fbbb \Bb^T \Ab^T).
\end{equation}
Replacing this expression in  \eqref{log-likelihood-SSCM-1} we get
\begin{align*}
	l(\Xbb|\Y, \Hb;\Ab, \Bb)  &=   - \frac{p}{2} \log |\Hb|   - \frac{n}{2} \log \Big|\frac{1}{n}(\Xbbb - \Fbbb \Bb^T \Ab^T )^T  (\Xbbb - \Fbbb \Bb^T \Ab^T )\Big|  -  \frac{1}{2} \tr( \ind_p) \nonumber
	\\ &=   - \frac{p}{2} \log |\Hb|  - \frac{n}{2} \log \Big|\frac{1}{n}(\Xbbb - \Fbbb \Bb^T \Ab^T )^T  (\Xbbb - \Fbbb \Bb^T \Ab^T )\Big|  -  \frac{p}{2}
\\ &=   - \frac{p}{2} \log |\Hb|  - \frac{n}{2} \log |\widetilde \Deltabf|  -  \frac{p}{2} . 
\end{align*}
Observe that maximizing $l(\Xbb|\Y, \Hb;\Ab, \Bb)$ is equivalent to minimizing $\log |\widetilde \Deltabf|$, which in turn is equivalent to minimizing $|\widetilde \Deltabf|$. Let $\widehat \Deltabf_{LS} = \frac{1}{n}(\Xbbb - \Fbbb \widehat{\Cb}_{LS}^T )^T (\Xbbb - \Fbbb \widehat{\Cb}_{LS}^T)$ with $\widehat \Cb_{LS}$ the full rank Least Squares estimator of model \eqref{model-last} with $\Cb=\Ab\Bb$. That is,  $\widehat \Cb_{LS} = \frac{1}{n}\widehat\Sigmabf_{\Xbbb,\Fbbb}\widehat\Sigmabf_{\Fbbb,\Fbbb}^{-1}$ with $\widehat \Sigmabf_{\Xbbb,\Fbbb} = \frac{1}{n} \Xbbb^T \Fbbb$ and $\widehat\Sigmabf_{\Fbbb,\Fbbb}  = \frac{1}{n} \Fbbb^T \Fbbb$, which are  the sample version of the covariance matrices $\Sigmabf_{\X, \f}$ and $\Sigmabf_{\f, \f}$, respectively. Following \cite{ReinselVelu98}, we can consider minimizing $|\widehat \Deltabf_{LS} \widetilde \Deltabf|$ instead of $|\widetilde \Deltabf|$  since $|\widehat \Deltabf_{LS}|$ is positive and constant with respect to $\Ab$ and $\Bb$. Consequently, by Theorem 2.2 of \cite{ReinselVelu98}, the estimators of  $\Ab$ and $\Bb$, are given by
\begin{equation}\label{AB-SSCM}
	\widehat{\Ab}=\widehat \Deltabf_{LS}^{-1/2}\widehat{\Vbf}_{(d)}\quad\text{and}\quad\widehat{\Bb}=\widehat{\Vbf}_{(d)}^{T}\widehat \Deltabf_{LS}^{1/2} \widehat\Sigmabf_{\Xbbb,\Fbbb}\widehat\Sigmabf_{\Fbbb,\Fbbb}^{-1},
\end{equation}
where $\widehat{\Vbf}_{(d)}=[\widehat{\Vbf}_{1},\ldots,\widehat{\Vbf}_{d}]$ are the eigenvectors that corresponds to the j-th largest eigenvalues $\lambda_j^2$ of
$\widehat \Sigmabf \doteq  \widehat \Deltabf_{LS}^{-1/2}\widehat \Sigmabf_{\Xbbb,\Fbbb} \widehat\Sigmabf_{\Fbbb,\Fbbb}^{-1} \widehat\Sigmabf_{\Fbbb,\Xbbb}\widehat \Deltabf_{LS}^{-1/2}$ where
$\widehat \Sigmabf_{\Xbbb,\Fbbb}$ and  $\widehat\Sigmabf_{\Fbbb,\Fbbb}$ were given above and $ \widehat\Sigmabf_{\Fbbb,\Xbbb} = \widehat \Sigmabf_{\Xbbb,\Fbbb}^T$.

Finally, with \eqref{AB-SSCM} in \eqref{mu-SSCM} and \eqref{Delta-SSCM} we have that
\begin{align}\label{muDeltaR-SSCM}
\widehat\mubf &=  (\Xbb^T - \widehat\Ab \widehat\Bb \Fbb^T)\Hb^{-1}\ones_n (\ones_n^T\Hb^{-1}\ones_n)^{-1}\\
\widehat \Deltabf &=  \frac{1}{n}(\Xbbb^T -  \widehat \Ab \widehat \Bb \Fbbb^T) (\Xbbb^T - \widehat \Ab \widehat \Bb \Fbbb^T)^T.
\end{align}
\end{proofprop}

\section{Proofs of Section \ref{secSEM} }\label{AppendixB}

\begin{proofprop}{SEM-estimators}
Given the matrix $\Wbf$ and the coefficient $\theta$, the next step is to estimate the mean $\mubf$. Taking derivative with respect to $\mubf$ in \eqref{log-likelihood-SEM}, except for some terms  independent of $\mubf$ we have that,
\begin{align*}
	\frac{\partial l(\Xbb|\Y, \theta;\mubf,\Ab, \Bb, \Deltabf )}{\partial \mubf}&= \\ &\hspace{-2cm} = \frac{\partial}{\partial \mubf}\left\{-  \frac{1}{2} \tr\Big(\Wbf_{\theta}(\Xbb -  \ones_n \mubf^T  - \Fbb \Bb^T \Ab^T ) \Deltabf^{-1} (\Xbb -  \ones_n \mubf^T - \Fbb \Bb^T \Ab^T)^T \Wbf_{\theta}\Big) \right\} \\ &\hspace{-2cm}= \Deltabf^{-1} (\Xbb^T -  \mubf \ones_n ^T - \Ab\Bb \Fbb^T  ) \Wbf_{\theta}^2 \ones_n,
\end{align*}
which is 0  if and only if 
\begin{equation}\label{mu-SEM}
\widetilde\mubf = (\Xbb^T - \Ab \Bb \Fbb^T){\Wbf}_{\theta}^2\ones_n (\ones_n^T{\Wbf}_{\theta}^2\ones_n)^{-1}.
\end{equation}
Therefore, in \eqref{log-likelihood-SEM} we have that
\begin{align*}
\Xbb -  \ones_n \widetilde \mubf^T - \Fbb\Bb^T \Ab^T & =\Xbb -  \ones_n 
\widetilde\mubf_{\Xbb}^T  - \widetilde\mubf_{\Fbb}^T \Bb^T \Ab^T  - \Fbb \Bb^T \Ab^T  \\ &= \Xbb -  (\ones_n^T{\Wbf}_{\theta}^2\ones_n)^{-1}  \ones_n  \ones_n^T {\Wbf}_{\theta}^2  \Xbb -  (\ones_n^T{\Wbf}_{\theta}^2\ones_n)^{-1} \ones_n  \ones_n^T {\Wbf}_{\theta}^2 \Fbb \Bb^T \Ab^T - \Fbb\Bb^T \Ab^T \\
	&= (\ind_n - (\ones_n^T{\Wbf}_{\theta}^2\ones_n)^{-1} \ones_n  \ones_n^T {\Wbf}_{\theta}^2)\Xbb - (\ind_n - (\ones_n^T{\Wbf}_{\theta}^2\ones_n)^{-1} \ones_n  \ones_n^T {\Wbf}_{\theta}^2) \Fbb\Bb^T \Ab^T \\ &\doteq   \Wbf^c_{\theta} \Xbb -  \Wbf^c_{\theta} \Fbb \Bb^T \Ab^T,
\end{align*}
where $\Wbf^c_{\theta}\doteq \ind_n -  (\ones_n^T{\Wbf}_{\theta}^2\ones_n)^{-1} \ones_n  \ones_n^T 
{\Wbf}_{\theta}^2$, 
Defining $\Xbbb$ and $\Fbbb$ as 
\[
\Xbbb = \Wbf_{\theta} \Wbf^c_{\theta} \Xbb\qquad \text{ and } \qquad \Fbbb = \Wbf_{\theta} \Wbf^c_{\theta} \Fbb,
\]
we have that, apart from irrelevant constants, the log-likelihhod \eqref{log-likelihood-SEM} becomes
\begin{align}\label{log-likelihood-SEM-1}
	l(\Xbb|\Y, \theta;\Ab, \Bb, \Deltabf)  &=  - \frac{n}{2} \log |\Deltabf| + p\log |\Wbf_{\theta}|  -  \frac{1}{2} \tr\Big( (\Xbbb - \Fbbb \Bb^T \Ab^T) \Deltabf^{-1} (\Xbbb- \Fbbb \Bb^T \Ab^T )^T  \Big).
\end{align}
Now, we take derivative with respect to $\Deltabf$  to get
\begin{align*}
	\frac{\partial l(\Xbb|\Y, \theta;\Ab, \Bb, \Deltabf )}{\partial \Deltabf^{-1}}&=  \frac{\partial}{\partial \Deltabf^{-1}} \left\{\frac{n}{2} \log |\Deltabf^{-1}| -  \frac{1}{2} \tr\Big((\Xbbb - \Fbbb \Bb^T \Ab^T ) \Deltabf^{-1} (\Xbbb- \Fbbb \Bb^T \Ab^T )^T \Big)\right\} \\ &= \frac{n}{2}\Deltabf -  \frac{1}{2} (\Xbbb - \Fbbb \Bb^T \Ab^T )^T (\Xbbb - \Fbbb \Bb^T \Ab^T )
\end{align*}
which is 0  if and only if 
\begin{equation}\label{Delta-SEM}
	\widetilde \Deltabf =  \frac{1}{n}(\Xbbb - \Fbbb \Bb^T \Ab^T )^T (\Xbbb - \Fbbb \Bb^T \Ab^T).
\end{equation}
Replacing this expression in \eqref{log-likelihood-SEM-1} we get
\begin{align*}
	l(\Xbb|\Y, \theta; \Ab, \Bb)  &=   - \frac{n}{2} \log \Big|\frac{1}{n}(\Xbbb - \Fbbb \Bb^T \Ab^T )^T  (\Xbbb - \Fbbb \Bb^T \Ab^T )\Big| + p\log |\Wbf_{\theta}|  -  \frac{1}{2} \tr \{\ind_p\}\nonumber
	\\ &=  - \frac{n}{2} \log \Big|\frac{1}{n}(\Xbbb - \Fbbb \Bb^T \Ab^T )^T  (\Xbbb - \Fbbb \Bb^T \Ab^T )\Big| + p\log |\Wbf_{\theta}|  -  \frac{p}{2} 	
	\\ &=  - \frac{n}{2} \log |\widetilde \Deltabf | + p\log |\Wbf_{\theta}|  -  \frac{p}{2} . 
\end{align*}

Again, as in the SSCM estimation, to  maximize $l(\Xbb|\Y, \theta;\Ab, \Bb)$ we we can consider minimizing $|\widehat \Deltabf_{LS} \widetilde \Deltabf|$ with $\widehat \Deltabf_{LS} = \frac{1}{n}(\Xbbb - \Fbbb \widehat{\Cb}_{LS}^T )^T (\Xbbb - \Fbbb \widehat{\Cb}_{LS}^T)$  and $\widehat \Cb_{LS}$ the full rank Least Squares estimator of model \eqref{model-last} with $\Cb=\Ab\Bb$. Then, from  \cite{ReinselVelu98}, the estimators of  $\Ab$ and $\Bb$ are given by
\begin{equation}\label{AB-SEM}
	\widehat{\Ab}=\widehat \Deltabf_{LS}^{-1/2}\widehat{\Vbf}_{(d)}\quad\text{and}\quad\widehat{\Bb}=\widehat{\Vbf}_{(d)}^{T}\widehat \Deltabf_{LS}^{1/2} \widehat\Sigmabf_{\Xbbb,\Fbbb}\widehat\Sigmabf_{\Fbbb,\Fbbb}^{-1},
\end{equation}
where $\widehat{\Vbf}_{(d)}=[\widehat{\Vbf}_{1},\ldots,\widehat{\Vbf}_{d}]$ are the eigenvectors that corresponds to the j-th largest eigenvalues $\lambda_j^2$ of
$\widehat \Sigmabf \doteq  \widehat \Deltabf_{LS}^{-1/2}\widehat \Sigmabf_{\Xbbb,\Fbbb} \widehat\Sigmabf_{\Fbbb,\Fbbb}^{-1} \widehat\Sigmabf_{\Fbbb,\Xbbb}\widehat \Deltabf_{LS}^{-1/2}$, with $\widehat \Sigmabf_{\Xbbb,\Fbbb}$ and  $\widehat\Sigmabf_{\Fbbb,\Fbbb}$ and $\widehat\Sigmabf_{\Fbbb,\Xbbb}$ as previously stated.  
Finally, with \eqref{AB-SEM} in \eqref{mu-SEM} and \eqref{Delta-SEM} we have that
\begin{align}\label{muDeltaR-SEM}
	\widehat\mubf &= (\Xbb^T -\widehat \Ab \widehat\Bb \Fbb^T){\Wbf}_{\theta}^2\ones_n (\ones_n^T{\Wbf}_{\theta}^2\ones_n)^{-1}\nonumber\\
	\widehat \Deltabf &=  \frac{1}{n}(\Xbbb^T -  \widehat \Ab \widehat \Bb \Fbbb^T) (\Xbbb^T - \widehat \Ab \widehat \Bb \Fbbb^T)^T. 		
\end{align}
\end{proofprop}

\end{document}